\RequirePackage{etoolbox}
\csdef{input@path}{{style/}{graphics/}}
\documentclass[MSNbibl,nameyear,dvips]{arxstspdf}
\usepackage{graphicx}
\usepackage{url,breakurl}
\usepackage{flushend}
\usepackage{stfloats}

\volume{29}
\issue{4}
\pubyear{2014}
\firstpage{559}
\lastpage{578}
\doi{10.1214/14-STS501} % kopijuoti is laisko
\docsubty{FLA}


\begin{document}
\begin{frontmatter}

\title{Causal Diagrams for Interference}%\thanksref{T1}
% kai straipsnis turi susijusiu diskusiju ir rejoinder'iu
%rejoinder at \relateddoi{r}{10.1214/00-STSXXXX}.}
\runtitle{Causal Diagrams for Interference}

\begin{aug}
\author[A]{\fnms{Elizabeth L.}~\snm{Ogburn}\corref{}\ead[label=e1]{eogburn@jhsph.edu}}%,
\and
\author[B]{\fnms{Tyler J.}~\snm{VanderWeele}\ead[label=e2]{tvanderw@hsph.harvard.edu}}
\runauthor{E. L. Ogburn and T. J. VanderWeele}

\affiliation{Johns Hopkins Bloomberg School of Public Health, Harvard
School of Public Health}

\address[A]{Elizabeth L. Ogburn is Assistant Professor, Department of
Biostatistics,
Johns Hopkins Bloomberg School of Public Health, 615 N. Wolfe Street,
Room E3620,
Baltimore, Maryland 21205, USA \printead{e1}.}
\address[B]{Tyler J. VanderWeele is Professor, Department of
Epidemiology, Harvard School of Public Health,
677 Huntington Avenue,
Kresge Building,
Boston, Massachusetts 02115, USA \printead{e2}.}
\end{aug}

% ABSTRACT
%
\begin{abstract}
The term ``interference'' has been
used to describe any setting in which one subject's
exposure may affect another subject's outcome. We
use causal diagrams to distinguish among three causal mechanisms that
give rise to interference. The first causal mechanism by which interference
can operate is a direct causal effect of one individual's treatment
on another individual's outcome; we call this \emph{direct interference}.
\emph{Interference by contagion} is present when one individual's
outcome may affect the outcomes of other individuals with whom he
comes into contact. Then giving treatment to the first individual
could have an indirect effect on others through the treated individual's
outcome. The third pathway by which interference may operate is
\emph{allocational
interference}. Treatment in this case allocates individuals to groups;
through interactions within a group, individuals may affect one another's
outcomes in any number of ways. In many settings, more than one type
of interference will be present simultaneously. The causal effects
of interest differ according to which types of interference are present,
as do the conditions under which causal effects are identifiable.
Using causal diagrams for interference, we describe these differences,
give criteria for the identification of important causal effects,
and discuss applications to infectious diseases.
\end{abstract}

% KEYWORDS
\end{frontmatter}

Traditionally, causal inference has relied on the assumption of
no interference, that is, the assumption that any subject's outcome
depends only on his own treatment and not on the treatment of any
other subject. This assumption is often implausible; for example, it
is violated when the outcome is an infectious disease and treating
one individual may have a protective effect on others in the population.
Recent work in statistics has focused on relaxing the assumption of
no interference (\citeauthor{graham2010measuring}, \citeyear
{graham2010measuring}; Halloran and Struchiner, \citeyear
{halloran1995causal};
\citeauthor{hudgens2008toward}, \citeyear{hudgens2008toward};
\citeauthor{manski2010identification}, \citeyear{manski2010identification};
\citeauthor{rosenbaum2007interference}, \citeyear
{rosenbaum2007interference}; Tchetgen Tchetgen and VanderWeele,
\citeyear{tchetgen2010causal};
\citeauthor{vansteelandt2007confounding}, \citeyear
{vansteelandt2007confounding}).
Much of this work has been motivated by the study of infectious diseases
(\citeauthor{halloran1995causal}, \citeyear{halloran1995causal};
\citeauthor{tchetgen2010causal}, \citeyear{tchetgen2010causal};
VanderWeele and Tchetgen Tchetgen,
\citeyear{vanderweele2011bounding,vanderweele2011effect};
\citeauthor{halloran2012}, \citeyear{halloran2012}).
Researchers have also explored the implications of interference on
residents of neighborhoods when some residents are given housing vouchers
to move (\citeauthor{sobel2006randomized}, \citeyear
{sobel2006randomized}) or when new resources are introduced
to the neighborhood (\citeauthor{vanderweele2010direct}, \citeyear
{vanderweele2010direct}). Others have written
about the interference that arises from assigning children to classrooms
and assigning classrooms to educational interventions
(\citeauthor{graham2010measuring}, \citeyear{graham2010measuring};
\citeauthor{hong2008causal}, \citeyear{hong2008causal}; \citeauthor
{vanderweele2013mediation}, \citeyear{vanderweele2013mediation}).
The rising prominence of social networks in public health research
underscores the need for methods that take into account the interconnections
among individuals' treatments and outcomes
(\citeauthor{christakis2007spread}, \citeyear{christakis2007spread};
Cohen-Cole and  Fletcher, \citeyear{cohen2008obesity};
\citeauthor{mulvaney2009obesity}, \citeyear{mulvaney2009obesity}).

Graphical models have shed light on the identification of causal effects
in many settings
(Dahlhaus and Eichler, \citeyear
{dahlhaus2003causality}; \citeauthor{didelez2010graphical}, \citeyear
{didelez2010graphical};
\citeauthor{freedman2004graphical}, \citeyear{freedman2004graphical};
\citeauthor{greenland1999causal}, \citeyear{greenland1999causal};
\citeauthor{pearl1995causal},
\citeyear{pearl1995causal,pearl1997graphical,pearl2000causality};
\citeauthor{robins2003semantics}, \citeyear{robins2003semantics};
Tian and Pearl, \citeyear{tian2002general};
\citeauthor{vansteelandt2007confounding}, \citeyear
{vansteelandt2007confounding})
but have not yet been applied to settings with interference. In this
paper, we describe how to draw causal diagrams representing the complex
interdependencies among individuals in the presence of interference,
and how to use those diagrams to determine what variables must be
measured in order to identify different causal effects of interest.
We review the literature on causal diagrams and identification of
causal effects in the absence of interference in Section~\ref{s1}, and recent
work on the estimation of causal effects in the presence of interference
in Section~\ref{s2}. In~Section~\ref{s3}, we discuss which covariates
must be measured
and controlled for in order to identify causal effects in the presence
of interference. Section~\ref{s4} introduces the three distinct types of
interference, provides causal diagrams to help explicate their structure,
and describes some of the causal effects we would wish to estimate
and the assumptions required to identify them. In Section~\ref{s5}, we use
the concepts introduced in Section~\ref{s4} to elucidate the nature of
interference
in social networks. Section~\ref{s6} concludes the paper.

%s1 #&#
\section{Review of Identification of Causal
Effects in the Absence of Interference}\label{s1}

Suppose that we wish to estimate the average causal effect of a treatment
$A$ on an outcome $Y$ from observational data on $n$ individuals
for whom we have also measured a vector of confounders $C$. For simplicity,
we will assume in this section and the next that $A$ is binary and
$Y$ is continuous, but our remarks apply equally to $A$ and $Y$
discrete or continuous. Under the assumptions of no interference and
a single version of treatment (we will not discuss the latter assumption
here; see \citeauthor{vanderweele2011causal}, \citeyear
{vanderweele2011causal}, for discussion), $Y_{i}(a), a=0,1$
is defined as the counterfactual outcome we would have observed if,
possibly contrary to fact, subject $i$ had received treatment $a$.
The average causal effect of $A$ on $Y$ is equal to $E [Y(1) ]-E [Y(0) ]$,
and it is identified under the three additional assumptions of consistency,
%
%e1 #&#
%
\begin{equation}
\label{consistency} Y_{i}(a)=Y_{i}\quad\mbox{if
}A_{i}=a,
\end{equation}
conditional exchangeability,
%
%e2 #&#
%
\begin{equation}\label{exchangeability}
Y_{i}(a)\amalg A_{i} |C_{i},
\end{equation}
and positivity,
%
%e3 #&#
%
\begin{eqnarray}\label{positivity}
&&P (A_{i}=a|C_{i}=c ) > 0\nonumber
\\
&&\hphantom{\hspace*{21pt}}\mbox{for all }a\mbox{ in the
support of }A\mbox{ and for all }c
\\
& & \hphantom{\hspace*{21pt}}\mbox{in the support of }C
\mbox{ such that }P(C=c)>0.
\nonumber
\end{eqnarray}
We refer the reader to \citet{Hernan2006estimating} for discussion
of these assumptions.

The conditional exchangeability assumption is sometimes referred to
as the ``no unmeasured confounding assumption.'' Identifying the
variables that must be included in $C$ can be assessed with the aid
of causal diagrams (e.g., \citeauthor{greenland1986identifiability},
\citeyear{greenland1986identifiability};
\citeauthor{greenland1999causal}, \citeyear{greenland1999causal};
\citeauthor{pearl2003statistics}, \citeyear{pearl2003statistics}).

\begin{figure}[t]

\includegraphics{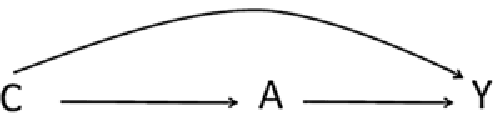}

\caption{}\label{f1}
\end{figure}

Causal diagrams, or causal directed acyclic graphs (DAGs) consist
of nodes, representing the variables in a study, and arrows, representing
causal effects. In a slight abuse of terminology, we will not distinguish
between nodes on a DAG and the variables they represent. A DAG is
a collection of nodes and arrows in which no variable is connected
to itself by a sequence of arrows aligned head-to-tail. A \emph{causal}
DAG is a DAG on which arrows represent causal effects and that includes
all common causes of any pair of variables on the graph. The causal
DAG in Figure~\ref{f1} represents the scenario in which the effect of $A$
on $Y$ is confounded by a single confounder $C$. The three arrows
encode the causal effects of $C$ on $A$, $C$ on $Y$, and $A$
on $Y$. We briefly introduce terminology and results for DAGs but
refer the reader to \citeauthor{pearl2000causality} (\citeyear
{pearl2000causality,pearl2003statistics})
for details and discussion. Recently, \citet{richardson2013single}
introduced a new class of causal diagrams called single world intervention
graphs (SWIGs). This work can be immediately and fruitfully applied
to the interference settings we discuss below; however, in the interest
of space we restrict our attention to DAGs.

%f1 #&#
%

A \emph{path} on a DAG is any unbroken, nonrepeating sequence of
arrows connecting one variable to another. A \emph{directed path}
(or a \emph{causal path} on a causal DAG) is a path that follows arrows
from tail to head. A variable $X$ is an \emph{ancestor} (or \emph{cause},
if the DAG is causal) of $Z$ if there is a directed path from $X$
to $Z$. Equivalently, $Z$ is a \emph{descendent} of $X$. If the
directed path from $X$ to $Z$ consists of a single arrow, then $X$
is a \emph{parent} of $Z$ and $Z$ is a \emph{child} of $X$. On
a causal DAG, we would say that $X$ has a \emph{direct effect} on
$Z$. If a path includes $X$, $W$ and $Z$ and if there are arrows
from both $X$ and $Z$ into $W$, then $W$ is a \emph{collider}
on the path. A collider is a path-specific concept. For example, in
Figure~\ref{f1}, $Y$ is a collider on one path from $C$ to $A$ (the path
$C\rightarrow Y\leftarrow A$) but not on another (the path
$C\rightarrow A\rightarrow Y$).
A path can be \textit{unblocked}, meaning roughly that information
can flow from one end to the other, or \textit{blocked}, meaning roughly
that the flow of information is interrupted at some point along the
path. If all paths between two variables are blocked, then the variables
are $d$-\textit{separated}, and if two variables are d-separated on a causal
DAG then they are statistically independent. A path is \textit{\emph{blocked}}
if there is a collider on the path such that neither the collider
itself nor any of its descendants is conditioned on. An unblocked
path can be blocked by conditioning on any noncollider along the
path. Two variables are d-separated by a set of variables if conditioning
on the variables in the set suffices to block all paths between them,
and if two variables are d-separated by a third variable or a set
of variables then they are independent conditional on the third variable
or set of variables (\citeauthor{pearl1995causal},
\citeyear{pearl1995causal,pearl2000causality}).

A \emph{backdoor path} from $X$ to $Z$ is one that begins with an
arrow pointing into, rather than out of, $X$. For example, the path
$A\leftarrow C\rightarrow Y$ in Figure~\ref{f1} is a backdoor path from
$A$ to $Y$. \citet{pearl1995causal} proved that conditioning on
a set of nondescendants of $A$ that block all backdoor paths from
$A$ to $Y$ suffices for exchangeability to hold for the effect of
$A$ on $Y$. This set need not be unique.

%f2 #&#
%
\begin{figure}[t]

\includegraphics{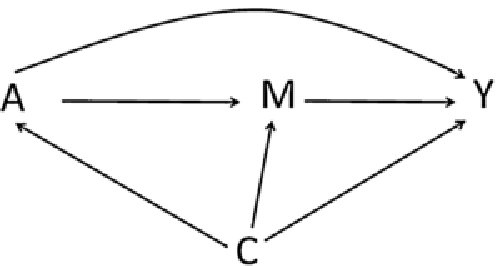}

\caption{}\label{f2}
\end{figure}

Identification of effects, other than total effects, often requires
assumptions beyond (\ref{consistency}), (\ref{exchangeability}),
and (\ref{positivity}). \emph{Path-specific effects} quantify the
causal effect of one variable on another via specific causal pathways.
Consider the DAG in Figure~\ref{f2}, which adds a mediator $M$ to the path
from $A$ to $Y$. Now there are two different causal pathways from
$A$ to $Y$, namely $A\rightarrow Y$ and $A\rightarrow M\rightarrow Y$.
Causal effects of the form $E [Y(a) ]-E [Y(a') ]$
capture all causal pathways from $A$ to $Y$ without distinguishing
among them, but we may be interested specifically in direct effects,
which bypass the mediator, and indirect effects, which go through
the mediator. Define $M_{i}(a)$ to be the counterfactual value we
would have observed for $M_{i}$ if $A_{i}$ had been set to $a$,
and $Y_{i}(a,m)$ to be the counterfactual value of $Y_{i}$ that
we would have observed if $M_{i}$ had been set to $m$ and $A_{i}$
to $a$. We make the additional consistency assumptions that $M_{i}(a)=M_{i}$
when $A_{i}=a$, that $Y_{i}(a,m)=Y_{i}$ when $A_{i}=a$ and $M_{i}=m$,
and that $Y_{i} (a,M_{i}(a) )=Y_{i}(a)$. Then the \emph{natural
direct effect} is defined as $E [Y (a,M(a) ) ]-E [Y (a',M(a) ) ]$;
it measures the expected change in $Y$ due to a change in $A$, holding
$M$ fixed at $M(a)$. A direct path from $X$ to $Z$ is said to
be \emph{deactivated} in a particular causal contrast if $X$ is set
to the same value in the counterfactual for $Z$ in both terms of
the contrast. A path is deactivated if any arrow on the path is deactivated.
In the natural direct effect, $A$ is set to the same value in the
counterfactual for $M$ in both terms of the contrast; therefore the
natural direct effect can be conceptualized as the effect of $A$
on $Y$ with the path $A\rightarrow M$ deactivated (\citeauthor
{pearl2001direct}, \citeyear{pearl2001direct}).
The \emph{natural indirect effect}, defined as $E [Y (a',M(a) ) ]-E [Y
(a',M(a') ) ]$,
measures the expected change in $Y$ when $A$ is held fixed but $M$
changes from $M(a)$ to $M(a')$. This is the effect of $A$ on $Y$
with the arrow from $A$ to $Y$ deactivated ($A$ is set to $a'$
in the counterfactual for $Y$ in both terms of the contrast). The
natural direct and indirect effects sum to the total effect of $A$
on $Y$: $E[Y(a)]-E[Y(a')] =E[Y(a,M(a))]-E[Y(a',M(a')]=
\{ E [Y (a,M(a) ) ]-E [Y (a',  M(a) ) ] \} + \{ E [Y (a',M(a) ) ]
-E [Y (a',\break M(a') ) ] \} $.
The \textit{controlled direct effect} of $A$ on $Y$, given by
$E[Y(a,m)]-E[Y(a',m)]$
fixes $M$ at a specific value $m$ and compares the counterfactual
outcomes under two different values of $A$. This is the effect of
$A$ on $Y$ with the path $M\rightarrow Y$ deactivated.

In order to identify the controlled direct effect, the following assumptions
are sufficient:
%
%e4 #&#
%
\begin{equation}\label{med no int 1}
Y_{i}(a,m)\amalg A_{i} |C_{i}
\end{equation}
and
%
%e5 #&#
%
\begin{equation}\label{med no int 2}
Y_{i}(a,m)\amalg M_{i} |A_{i},C_{i}.
\end{equation}
These correspond respectively to the absence of unblocked backdoor
paths from $A_{i}$ to $Y_{i}$ (except possibly through $M_{i}$)
conditional on $C_{i}$ and from $M_{i}$ to $Y_{i}$ conditional
on $A_{i}$ and $C_{i}$. \citet{avin2005identifiability} proved
that in most settings the following is a necessary assumption for
the identification of the average natural direct and indirect effects
of $A_{i}$ on $Y_{i}$ mediated by $M_{i}$: there is no variable
$W_{i}$ such that (i) there is an activated directed path from $A_{i}$
to $W_{i}$, (ii) there is a deactivated directed path from $W_{i}$
to $Y_{i}$ and (iii) there is an activated directed path from $W_{i}$
to $Y_{i}$. A variable that satisfies conditions (i), (ii) and (iii)
is known as a \emph{recanting witness}, and we call the assumption
of no variable satisfying these conditions the \emph{recanting witness
criterion}. In the context of natural direct and indirect effects,
the recanting witness criterion is met if there is no confounder of
the mediator--outcome relation that is caused by treatment, or
(\citeauthor{pearl2001direct}, \citeyear{pearl2001direct})
%
%e6 #&#
%
\begin{equation}\label{med no int 4}
Y_{i}(a,m)\amalg M_{i}\bigl(a'
\bigr)|C_{i}.
\end{equation}
Assumptions (\ref{med no int 2}), (\ref{med no int 1}), (\ref{med no
int 4})
and
%
%e7 #&#
%
\begin{equation}\label{med no int 3}
M_{i}(a)\amalg A_{i} |C_{i},
\end{equation}
that is, the absence of unblocked backdoor paths from $A_{i}$ to $M_{i}$
conditional on $C_{i}$, suffice to identify the natural direct and
indirect effects.

%s2 #&#
\section{Review of Identification of Causal Effects in the Presence of
Interference}\label{s2}

Interference is present when one subject's outcome may depend on other
subjects' treatments (\citeauthor{rosenbaum2007interference},
\citeyear{rosenbaum2007interference}). It is often
reasonable to make a \emph{partial interference} assumption that interference
can only occur within subgroups or \emph{blocks} of subjects. This
may be justified if the blocks are separated by time or
space (\citeauthor{hudgens2008toward}, \citeyear{hudgens2008toward};
\citeauthor{sobel2006randomized}, \citeyear{sobel2006randomized};
Tchetgen Tchetgen and VanderWeele, \citeyear{tchetgen2010causal};
\citeauthor{rosenbaum2007interference}, \citeyear{rosenbaum2007interference}).
Under interference, $Y_{i}(a)$ is not well-defined, since the value
of $Y$ that would have been observed for subject $i$ had he received
treatment $a$ may depend on the treatments received by other subjects.
We define counterfactual notation for interference following
\citet{hudgens2008toward},
\citet{tchetgen2010causal}, \citet{rubin1990application}
and \citet{halloran1995causal}.
Suppose that $n$ individuals fall into $N$ blocks, indexed by $k$,
with $m=n/N$ individuals in each block. If $N=1$, so that interference
may occur between any two subjects in the population, then we say
that there is full interference. If $N=n$, then an individual's treatment
can only affect his own outcome and there is no interference. Let
$\mathbf{A}_{k}\equiv(A_{k1},\ldots,A_{km} )$ be the vector
of treatment assignments for individuals in block $k$ and let $\mathbf{a}_{k}$
denote an $m$-dimensional vector in the support of $\mathbf{A}_{k}$.
Let $\mathbf{Y}_{k}\equiv(Y_{k1},\ldots,Y_{km} )$ and
$\mathbf{C}_{k}\equiv(C_{k1},\ldots,C_{km} )$ be the vector
of outcomes and array of covariates, respectively, for individuals
in block $k$. In what follows, we reserve boldface letters for vectors
or arrays of length $m$ in which the $i$th entry corresponds to
the $i$th individual in block $k$, and we omit the subscript
$k$ when taking expectations over blocks. Define $Y_{ki}(\mathbf{a}_{k})$
to be the counterfactual outcome that we would have observed for individual
$i$ in block $k$ under an intervention that set $\mathbf{A}_{k}$
to $\mathbf{a}_{k}$. Following \citet{tchetgen2010causal} we
replace assumption (1) above with a new assumption of consistency
under interference:
%
%e8 #&#
%
\begin{equation}\label{consistency interference}
Y_{ki}(\mathbf{a}_{k})=Y_{ki}\quad\mbox{when }
\mathbf{A}_{k}=\mathbf{a}_{k} .
\end{equation}
We also require modified positivity and exchangeability assumptions
in order to identify causal effects under interference: we assume
that we have measured a set of pretreatment covariates $C$ for each
individual such that (Tchetgen Tchetgen and VanderWeele, \citeyear
{tchetgen2010causal})
%
%e9 #&#
%
\begin{equation}\label{exchangeability interference}
Y_{ki}(\mathbf{a}_{k})\amalg\mathbf{A}_{k} |
\mathbf{C}_{k}
\end{equation}
and
%
%e10 #&#
%
\begin{eqnarray}\label{positivity interference}
&&P (\mathbf{A}_{k}=\mathbf{a}_{k}|\mathbf{C}_{k}=
\mathbf{c}_{k} )>0 \nonumber
\\
&& \hspace*{44pt}\mbox{for all }\mathbf{a}_{k}\mbox{
in the support of }\mathbf{A}_{k} \hspace*{-101pt}
\\
&& \hspace*{44pt}\mbox{and for all $\mathbf{c}_{k}$ in the support
of }
\mathbf{C}_{k}. \nonumber\hspace*{-10pt}
\end{eqnarray}

These assumptions suffice to identify the expectations of counterfactuals
of the form $Y_{ki}(\mathbf{a}_{k})$ whenever $\mathbf{a}_{k}$
is an instance of a well-defined intervention $\mathbf{a}$ and,
therefore, to identify causal effects that are contrasts of such expectations.
An intervention will be well-defined if it uniquely determines which
subjects in block receive treatment. Well-defined interventions are
possible, for example, if all blocks are of the same size, if the
individuals in each block are distinguishable from one another, and
if the individuals are ordered in the same way across blocks. Suppose
interference occurs within blocks comprised of a father (subject 1),
a mother (subject 2) and a child (subject 3). Then an intervention
$(1, 0, 1)$ indicates that the father and child receive treatment
but the mother does not. If the blocks are of different sizes or if
there is no natural way to distinguish among the individuals in each
block, some interventions may be well-defined under assumptions that
the effects of treatment are the same for different members of the
block and do not depend on the size of the block, for example, the intervention
that assigns treatment to every individual in every block. We assume
throughout that all interventions are well-defined. For simplicity,
we assume that the blocks are of the same size and that there is a
natural ordering of the subjects in each block, but most of our comments
and results extend to more general settings. (In the absence of well-defined
interventions, some causal effects can still be defined, identified
and estimated under two-stage randomization; see \citeauthor
{hudgens2008toward}, \citeyear{hudgens2008toward};
\citeauthor{vanderweele2011components}, \citeyear{vanderweele2011components};
\citeauthor{vanderweele2011effect}, \citeyear{vanderweele2011effect};
Halloran and Hudgens, \citeyear{halloran2012}.)

In Section~\ref{s3}, we use graphical models to determine which variables
must be included in the conditioning set in order for exchangeability
to hold. This gives the identification criteria under interference
for causal effects that are contrasts of expectations of counterfactuals
of the form $Y_{ki}(\mathbf{a}_{k})$. Sometimes we may wish to
identify path-specific effects; these require additional assumptions
for identification that we discuss below.

In this paper, we focus on identification, rather than estimation,
of causal effects. We merely note here that, for the purposes of estimation
and inference, the effective sample size is $N$ and thus observation
of multiple blocks may be required.

%s2.1 #&#
\subsection{Causal Effects of $\mathbf{A}$ on $Y$}\label{s2.1}

Although recognition of the fact that interference may occur in certain
settings dates at least as far back as \citet{ross1916application},
it is only recently that progress has been made on identifying causal
effects in the presence of interference. \citet{halloran1995causal}
defined four effects that do not depend on understanding the mechanisms
underlying interference and that are identifiable under assumptions
(\ref{consistency interference}), (\ref{exchangeability interference})
and (\ref{positivity interference}).

The overall effect of intervention $\mathbf{a}$ compared to intervention
$\mathbf{a}'$ on subject $i$ is defined as $\mathit{OE}_{i}(\mathbf
{a},\mathbf{a}')\equiv E[Y_{i}(\mathbf{a})]-E[Y_{i}(\mathbf{a}')]$.
We use the index $i$ to indicate that the expectations do not average
over individuals within a block but rather over blocks for a particular
individual $i$. For example, if the blocks are comprised of a father
(subject 1), a mother (subject 2) and a child (subject 3), then
$\mathit{OE}_{3}(\mathbf{a},\mathbf{a}')=E[Y_{3}(\mathbf
{a})]-E[Y_{3}(\mathbf{a}')]$
is the overall effect on a child of intervention $\mathbf{a}$
compared to intervention $\mathbf{a}'$. The average overall effect
$\mathit{OE}(\mathbf{a},\mathbf{a}')\equiv E[Y(\mathbf
{a})]-E[Y(\mathbf{a}')]$,
where $E[Y(\mathbf{a})]\equiv\frac{1}{m}\sum_{i=1}^{m}E
[Y_{i}(\mathbf{a}) ]$,
averages over the empirical mean of the counterfactual outcomes for
each block. The unit-level effect of treatment on subject $i$ fixes
the treatment assignments for all subjects in each block except~$i$,
and compares the counterfactual outcomes for subject $i$ under two
different treatment assignments. Let
$\mathbf{a}_{k,-i}=(a_{k,1},\ldots,a_{k,i-1},a_{k,i+1},\ldots, a_{k,m})$
be a vector of length $m-1$ of treatment values for all subjects
in block $k$ except for subject $i$. Then $\mathit{UE}_{i}(\mathbf
{a};\tilde{a},\bar{a})\equiv E[Y_{i}(\mathbf{a}_{-i},\tilde{a})]-E[Y_{i}
(\mathbf{a} _{-i},\bar{a})]$,
where $Y_{ki}(\mathbf{a}_{k,-i},\tilde{a})$ is subject $i$'s
counterfactual outcome under the intervention in which the subjects
in block $k$ except for subject $i$ receive treatments $\mathbf{a}_{k,-i}$
and subject $i$ receives treatment $\tilde{a}$. The spillover effect
of intervention $\mathbf{a}$ compared to intervention $\mathbf{a}'$
on subject $i$ fixes $i$'s treatment level and compares his counterfactual
outcomes under the two different interventions. That is, $\mathit
{SE}_{i}(\mathbf{a},\mathbf{a}';\tilde{a})\equiv E[Y_{i}(\mathbf
{a}_{-i},\tilde{a})]-E[Y_{i}(\mathbf{a}_{-i}',\tilde{a})]$.
(The unit-level effect is often referred to as the direct effect and
the spillover effect as the indirect effect of an intervention, but
in order to avoid confusion with the direct effect for DAGs defined
in Section~\ref{s2} and the natural direct and indirect effects
defined in
Section~\ref{s2.2}, we will use different terminology. See
\citeauthor{tchetgen2010causal}, \citeyear{tchetgen2010causal},
and Vanderweele, Tchetgen and Halloran, \citeyear
{vanderweele2011components}, for further discussion of
terminology.) We can also average these effects over individuals within
a block. The average unit-level effect is $\mathit{UE}(\mathbf
{a};\tilde{a},\bar{a})\equiv E[Y(\mathbf{a}_{-},\tilde
{a})]-E[Y(\mathbf{a}_{-},\bar{a})]$
and the average spillover effect is $\mathit{SE}(\mathbf{a},\mathbf
{a}';\tilde{a})\equiv E[Y(\mathbf{a}_{-},\tilde{a})]-E[Y(\mathbf
{a}_{-}',\tilde{a})]$,
where $E [Y(\mathbf{a}_{-},\tilde{a}) ]\equiv\frac{1}{m}\sum
_{i=1}^{m}E [Y_{i}(\mathbf{a}_{-i},\tilde{a}) ]$.
The total effect compares an individual's counterfactual at one treatment
level in a block that receives one intervention to his counterfactual
at a different treatment level in a block that receives the another
intervention: $\mathit{TE}_{i}(\mathbf{a},\mathbf{a}';\tilde
{a},\bar{a})\equiv E[Y_{i}(\mathbf{a}_{-i},\tilde
{a})]-E[Y_{i}(\mathbf{a}_{-i}',\bar{a})]$.
The average total effect is defined analogously to the other average
effects above. \citet{hudgens2008toward} showed that the total effect
can be decomposed into a sum of unit-level and spillover effects:
$\mathit{TE}_{i}(\mathbf{a},\mathbf{a}';\tilde{a},\bar{a})=
E[Y_{i}(\mathbf{a}_{-i},\tilde{a})]-E[Y_{i}(\mathbf{a}_{-i},\bar{a})]
+  E[Y_{i}(\mathbf{a}_{-i},\bar{a})]-E[Y_{i}(\mathbf
{a}_{-i}',\bar{a})]
=\mathit{DE}_{i}(\mathbf{a};\break\tilde{a},\bar{a})+ \mathit
{IE}_{i}(\mathbf{a},\mathbf{a}';
\bar{a})$.

\citet{sobel2006randomized}, \citet
{hudgens2008toward}, \citet{vanderweele2011effect}
and \citet{tchetgen2010causal} proposed ways to estimate and extend
unit-level, spillover, total and overall effects. We will not discuss
these extensions here except to note that they require the same three
identifying assumptions (\ref{consistency interference}), (\ref
{exchangeability interference})
and (\ref{positivity interference}).

%s2.2 #&#
\subsection{Path Specific Effects of $\mathbf{A}$ on $Y$}\label{s2.2}

In Sections~\ref{s4.2} and \ref{s4.3},
we describe path-specific effects that may
be of interest in certain interference contexts; here we review and
extend the literature on path-specific effects under interference.
Let $\mathbf{M}_{k}\equiv(M_{k1},\ldots,M_{km} )$ be a
vector of variables that may lie on a causal pathway from $\mathbf{A}_{k}$
to $Y_{ki}$. \citet{vanderweele2010direct} provided identifying
assumptions and expressions for mediated effects with cluster-level
treatments. These effects are applicable to the context of partial
interference where there is interference by the mediator but not by
the treatment ($M_{ki}$ may have an effect on $Y_{kj}$ for $i\neq j$,
but $\mathbf{A}_{k}$ is set at the cluster level, and thus is the
same for all individuals in block $k$). We adapt them here to accommodate
interference by the treatment in addition to the mediator. We make
the consistency assumptions that $\mathbf{M}_{k}(\mathbf
{a}_{k})=\mathbf{M}_{k}$
when $\mathbf{A}_{k}=\mathbf{a}_{k}$, that $Y_{kj}(\mathbf
{a}_{k},\mathbf{m}_{k})=Y_{kj}$
when $\mathbf{A}_{k}=\mathbf{a}_{k}$ and $\mathbf{M}_{k}=\mathbf{m}_{k}$,
and that $Y_{kj}(\mathbf{a}_{k},\mathbf{M}_{k}(\mathbf
{a}_{k}))=Y_{kj}(\mathbf{a}_{k})$.
The expected controlled direct effect of a block-level treatment
$\mathbf{A}_{k}$
on individual $i$'s outcome, not through $\mathbf{M}_{k}$, is
defined as $E [Y_{i} (\mathbf{a},\mathbf{m} ) ]-E [Y_{i} (\mathbf
{a}',\mathbf{m} ) ]$;
it measures the expected change in $Y_{ki}$ due to a change in $\mathbf
{A}_{k}$,
intervening to set $\mathbf{M}_{k}$ to $\mathbf{m}_{k}$.
The expected natural direct effect is $E [Y_{i} (\mathbf{a},\mathbf
{M}(\mathbf{a}) ) ]-E [Y_{i} (\mathbf{a}',\mathbf{M}(\mathbf{a}) ) ]$;
it measures the expected change in $Y_{ki}$ due to a change in $\mathbf
{A}_{k}$,
holding $\mathbf{M}_{k}$ fixed at $\mathbf{M}_{k}(\mathbf{a}_{k})$.
The expected natural indirect effect of $\mathbf{A}_{k}$ on $Y_{ki}$
through $\mathbf{M}_{k}$, given by $E [Y_{i} (\mathbf{a}',\mathbf
{M}(\mathbf{a}) ) ]-E [Y_{i} (\mathbf{a}',\mathbf{M}(\mathbf{a}') ) ]$,
measures the expected change in $Y_{ki}$ when $\mathbf{A}_{k}$
is fixed but $\mathbf{M}_{k}$ changes from $\mathbf{M}_{k}(\mathbf{a}_{k})$
to $\mathbf{M}_{k}(\mathbf{a}_{k}')$. Average controlled
direct, natural direct, and natural indirect effects are defined similarly
to the average effects in Section~\ref{s2.1}: we average the counterfactuals
within each block before taking the expectations over blocks. These
natural direct and indirect effects are identifiable under the following
four assumptions:
%
%e11 #&#
%e12 #&#
%e13 #&#
%
\begin{eqnarray}
\label{eq:med1}Y_{ki} (\mathbf{a}_{k},\mathbf{m}_{k} ) &\amalg&
\mathbf{A}_{k} |\mathbf{C}_{k},
\\
\label{eq:med2}Y_{ki} (\mathbf{a}_{k},\mathbf{m}_{k} ) &\amalg&
\mathbf{M}_{k} |\mathbf{A}_{k},\mathbf{C}_{k},
\\
\label{med3}\mathbf{M}_{k} (\mathbf{a}_{k} ) &\amalg&
\mathbf{A}_{k} |\mathbf{C}_{k}
\end{eqnarray}
and
%
%e14 #&#
%
\begin{equation}\label{med4}
Y_{ki} (\mathbf{a}_{k},\mathbf{m}_{k} ) \amalg
\mathbf{M}_{k} \bigl(\mathbf{a}_{k}' \bigr)|
\mathbf{C}_{k}.
\end{equation}
Assumptions (\ref{eq:med1}), (\ref{eq:med2}) and (\ref{med3})
correspond, respectively, to the absence of unblocked backdoor paths
from $\mathbf{A}_{k}$ to $Y_{ki}$ (except possibly through $\mathbf{M}_{k}$)
conditional on $\mathbf{C}_{k}$, from $\mathbf{M}_{k}$ to
$Y_{ki}$ conditional on $\mathbf{A}_{k}$ and $\mathbf{C}_{k}$,
and from $\mathbf{A}_{k}$ to $\mathbf{M}_{k}$ conditional
on $\mathbf{C}_{k}$. Assumption (\ref{med4}), similar to (\ref{med
no int 4}),
corresponds to the recanting witness criterion. Under these assumptions,
counterfactual expectations of the form $E [Y_{i} (\mathbf{a}',\mathbf
{M}(\mathbf{a}) ) ]$
are identified and, therefore, so are the natural direct and indirect
effects, which are contrasts of such expectations. Specifically, $E
[Y_{i} (\mathbf{a}',\mathbf{M}(\mathbf{a}) ) ]$
is identified by
\begin{eqnarray*}
&&\sum_{\mathbf{c}}\sum_{\mathbf{m}}E
\bigl[Y_{i}|\mathbf{A}=\mathbf{a}',\mathbf{M}=\mathbf{m},
\mathbf{C}=\mathbf{c} \bigr]
\\
&&\hphantom{\sum_{\mathbf{c}}\sum_{\mathbf{m}}}{}\cdot P (\mathbf
{M}=\mathbf{m}|\mathbf{A}=\mathbf{a},
\mathbf{C}=\mathbf{c} )P (\mathbf{C}=\mathbf{c} ).
\end{eqnarray*}
Assumptions (\ref{eq:med1}) and (\ref{eq:med2}) suffice to identify
the controlled direct effect.

%s3 #&#
\section{Covariate Control}\label{s3}

\begin{figure}[t]

\includegraphics{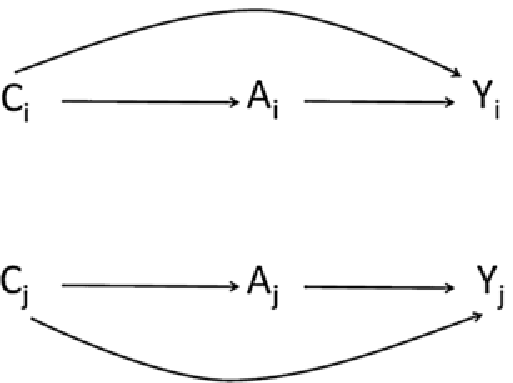}

\caption{}\label{f3}
\end{figure}

Although the subscripts are usually suppressed, under the assumption
of no interference the standard DAG for the effect of a treatment
$A$ on an outcome $Y$ with confounders $C$ is drawn to show the
relationships among $Y_{i}$, $A_{i}$ and $C_{i}$ for subject $i$.
Under interference, however, it is not sufficient to consider causal
pathways at the individual level; a causal DAG must depict an entire
block. For simplicity, we will focus on blocks of the smallest size
that preserves the essential structure of interference, which for
our purposes will be two or three. The principles extend to groups
of any size, but the DAGs become considerably more complex as the
blocks grow. The DAG for the effect of $A$ on $Y$ in a group of
size two with no interference is depicted in Figure~\ref{f3}. In what follows,
we represent a single block of subjects on each DAG, and we therefore
suppress the subscript $k$ indicating membership in block $k$.

%f3 #&#
%

%f4 #&#
%
\begin{figure}[t]

\includegraphics{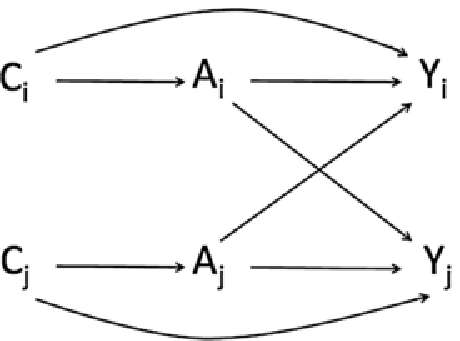}

\caption{}\label{f4}
\end{figure}

%f5 #&#
%
\begin{figure*}[b]
\begin{tabular}{ccc}%

\includegraphics{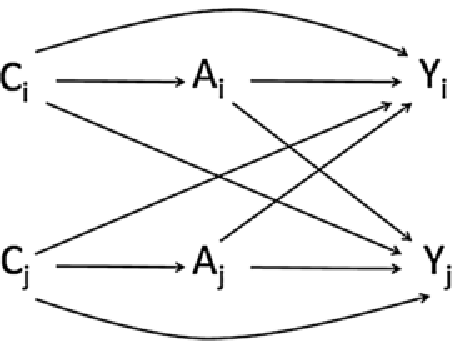} &\includegraphics{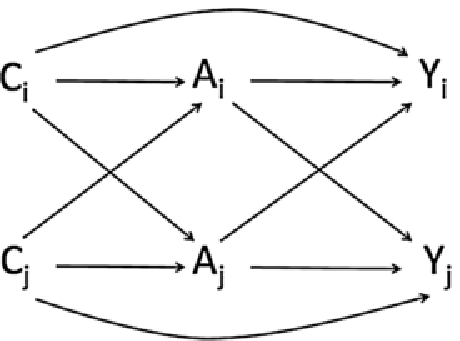}&\includegraphics{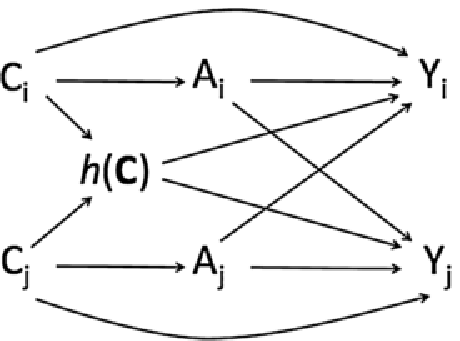}
\\[2pt]
(a)&(b)& (c)
\\[6pt]

\includegraphics{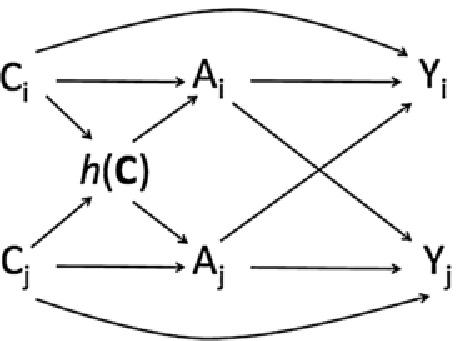} &\includegraphics{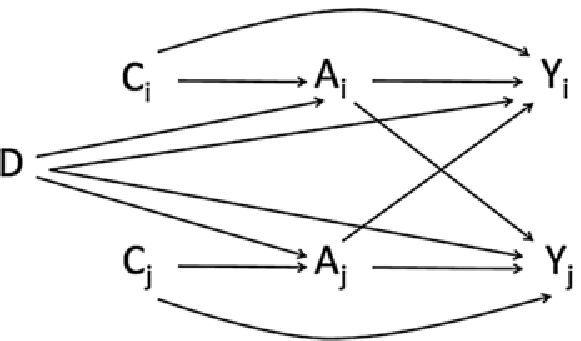}&\includegraphics{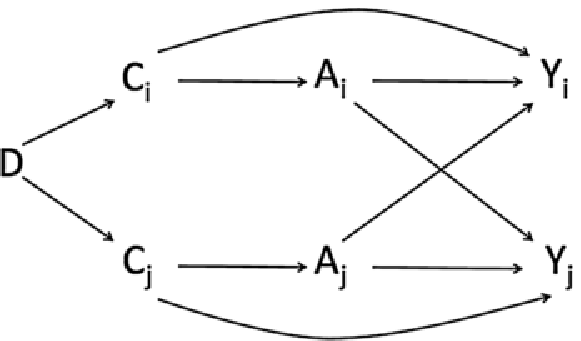}
\\[2pt]
(d)&(e)&(f)
\\[6pt]
\multicolumn{3}{c}{
\includegraphics{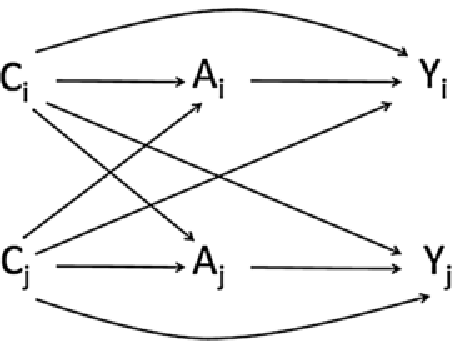}
}
\\[2pt]
\multicolumn{3}{c}{(g)}
\end{tabular}
\caption{}\label{f5}
\end{figure*}

Interference can be represented by a DAG like the one given in
Figure~\ref{f4}.
The arrows from $A_{i}$ to $Y_{j}$ for $i\neq j$ represent the
effect that one individual's treatment has on another's outcome. This
representation suffices whenever contrasts of counterfactuals of the
form $Y(\mathbf{a})$, such as the effects described in Section~\ref{s2.1},
are the only effects of interest. However, as we will see below,
when contagion or allocational interference are present, such a diagram
does not represent information about how the effect of $A_{i}$ on
$Y_{j}$ operates. In Section~\ref{s4}, we describe how to represent this
additional information on a DAG. We describe covariate control in
the general cases depicted in the DAGs in Figures~\ref{f4} and \ref
{f5} before moving
on in Section~\ref{s4} to tease apart the structures that make direct
interference,
interference by contagion, and allocational interference distinct.
The principles of covariate control in the presence of interference
are straightforward: like in the case of no interference, they follow
from the fact that all backdoor paths from treatment to outcome must
be blocked by a measured set of covariates. However, without taking
the time to draw the operative causal DAG with interference it is
easy to make mistakes, like controlling only for individual-level
covariates when block-level covariates are necessary to identify the
causal effect of interest. Below we will consider a number of different
settings and causal structures, and discuss in each whether control
for only an individual's covariates suffices to identify causal effects
or whether control for the covariates of the entire block (or of some
summary) is needed.

If the individuals in the block share no common causes of $A$ or
$Y$, as in the DAG in Figure~\ref{f4}, then $C_{i}$ suffices to block the
backdoor paths from $A_{i}$ to $Y_{i}$ and from $A_{j}$ to $Y_{i}$
and, therefore, exchangeability for the effect of $\mathbf{A}$
on $Y_{i}$ holds conditional on $C_{i}$. That is,
$ Y_{i}(a_{i},a_{j})\amalg\mathbf{A} |C_{i}$
for all $i$. If $C_{j}$ is a direct cause of $Y_{i}$ for $i\neq j$,
as in Figure~\ref{f5}(a), then exchangeability for the effect of
$\mathbf{A}$
on $Y_{i}$ necessitates block- and not just individual-level covariates.
Even if each individual's treatment is randomized conditional on his
own covariates (this corresponds to the absence of arrows $C_{j}$
to $A_{i}$ for $i\neq j$ on the DAG), there is still a backdoor
path from $A_{j}$ to $Y_{i}$ via $C_{j}$ and, somewhat counterintuitively,
it is necessary to control for $C_{j}$ in addition to $C_{i}$ in
any model for the effect of $
\mathbf{A}$ on $Y_{i}$. On
the other hand, if $C_{j}$ directly affects $A_{i}$ but not $Y_{i}$
for $i\neq j$, as in Figure~\ref{f5}(b), then for exchangeability for the
effect of $\mathbf{A}$ on $Y_{i}$ it suffices to condition only
on $C_{i}$. If, in addition to $C_{i}$, a function $h(\mathbf{C})$
of the vector of covariates influences outcome (Figure~\ref{f5}(c);
for example,
the mean value of $\mathbf{C}$ for the block), then $C_{i}$
and either $h(\mathbf{C})$ or $C_{j}$ need to be conditioned
on in order to achieve exchangeability for the effect of $\mathbf{A}$
on $Y_{i}$. If $h(\mathbf{C})$ only influences treatment assignment
(Figure~\ref{f5}(d)), then only
$C_{i}$ must be conditioned on. If a block-level characteristic $D$ is
a common
cause of $A$ and $Y$ (Figure~\ref{f5}(e)), then $C_i$ and $D$ must be
conditioned on in order to achieve exchangeability for the
effect of $\mathbf{A}$ on $Y_i$. If $C_{i}$ and $C_{j}$ share a
common cause (Figure~\ref{f5}(f)), then exchangeability for the effect
of $\mathbf{A}$
on $Y_{i}$ holds conditional on~$C_{i}$.

Even in the absence of interference for the effect of $\mathbf{A}$
on $\mathbf{Y}$, there are scenarios in which individual-level
covariates do not suffice to control for the effect of an individual's
treatment on his own outcome. For example, the DAG in Figure~\ref
{f5}(g) depicts
a scenario in which one individual's covariates affect another individual's
treatment and outcome (represented by the arrows $C_{i}\rightarrow A_{j}$
and $C_{i}\rightarrow Y_{j}$), but there is no effect of one individual's
treatment on another's outcome (no directed path from $A_{i}$ to
$Y_{j}$). In other words, there is interference for the effect of
$\mathbf{C}$ on $\mathbf{Y}$ but not for the effect of $\mathbf{A}$
on $\mathbf{Y}$. \citet{vansteelandt2007confounding} noted that
in this setting it is necessary to condition on $\mathbf{C}$
to achieve exchangeability for the effect of $A_{i}$ on $Y_{i}$.

Consider the DAG in Figure~\ref{f4}, but now suppose that $\mathbf{C}$
is unobserved. As we discussed above, $C_{i}$ is a confounder of
the effect of $A_{i}$ on $Y_{i}$, but in this DAG the effect of
$A_{j}$ on $Y_{i}$ is unconfounded (there is no backdoor path from
$A_{j}$ to $Y_{i}$). If a researcher hypothesizes that the DAG in
Figure~\ref{f4} represents the underlying causal structure in a particular
setting but he does not have access to data on the confounders $\mathbf{C}$,
then the effect of $\mathbf{A}$ on $Y_{i}$ is not identified.
However, the unconfounded effect of $A_{j}$ on $Y_{i}$ is identified
by $E[Y_{i}|A_{j}=a_{j}]$. This quantity has an interpretation in
the interference setting as the weighted average of expected counterfactuals
within strata of $C$.
\begin{eqnarray*}
&&E[Y_{i}|A_{j}=a_{j}]
\\
&&\quad= \sum
_{a_{i}}\sum_{c}E[Y_{i}|A_{i}=a_{i},A_{j}=a_{j},C_{i}=c]
\\
&&\hphantom{\quad= \sum
_{a_{i}}\sum_{c}}{}\cdot P(A_{i}=a_{i}|C_{i}=c)P(C_{i}=c)
\\
&&\quad= \sum_{a_{i}}\sum_{c}E
\bigl[Y_{i}(a_{i},a_{j})|A_{i}=a_{i},A_{j}=a_{j},C_{i}=c
\bigr]
\\
&&\hphantom{\quad= \sum
_{a_{i}}\sum_{c}}{}\cdot P(A_{i}=a_{i}|C_{i}=c)P(C_{i}=c)
\\
&&\quad= \sum_{a_{i}}\sum_{c}E
\bigl[Y_{i}(a_{i},a_{j})|C_{i}=c
\bigr]\\
&&\hphantom{\quad= \sum
_{a_{i}}\sum_{c}}{}\cdot P(A_{i}=a_{i}|C_{i}=c)P(C_{i}=c),
\end{eqnarray*}
where the first equality relies on the facts that $A_{i}\amalg A_{j}|C_{i}$
and $C_{i}\amalg A_{j}$, the second relies on consistency, and the
third on conditional exchangeability. Alternatively, this quantity
has the interpretation of $E[Y_{i}(a_{j})]$ in an experiment where
$Y_{i}$ is considered to be the only outcome, $A_{j}$ is the treatment
of interest and is intervened on, and $A_{i}$ is randomly assigned
according to the actual distribution $P(A_{i}|C_{i})$ in the population.

The hypothetical experiment described above points toward a possible
strategy for estimating the effect of one component of $\mathbf{A}$
on $Y_{i}$ when the confounders of the effect of $\mathbf{A}$
on $Y_{i}$ are not fully observed. The researcher can analyze each
block of subjects as a single observation, with a single treatment
and outcome. This strategy discards data on others' treatments and
outcomes but may allow for progress even if the full set of covariates
needed to identify $Y_{i}(\mathbf{a})$ are not observed.

In some of the DAGs in Figure~\ref{f5}, identification of the effect
of $A_{j}$
on $Y_{i}$ requires fewer covariates than the effect of $\mathbf{A}$
on $Y_{i}$. In Figure~\ref{f5}(c), $C_{j}$ suffices to control for confounding
of the effect of $A_{j}$ on $Y_{i}$ even though it does not suffice
for the effect of $\mathbf{A}$ on $Y_{i}$. For the DAG in Figure~\ref{f5}(e),
$D$ suffices to control for confounding of the effect of $A_j$ on $Y_i$.

In some cases, we can identify the effect of $A_{i}$ on $Y_{i}$ with
fewer covariates than are required to identify the effect of $\mathbf{A}$
on $Y_{i}$. In Figures~\ref{f5}(a) and~\ref{f5}(c), we can identify
the effect
of $A_{i}$ on $Y_{i}$ if only $C_{i}$ is observed, even though
we cannot identify the effect of $\mathbf{A}$ on $Y_{i}$ without
also observing $C_{j}$ (or $h(\mathbf{C})$). In~the
\hyperref[app]{Appendix}, we give the identifying expressions for
these effects.

For the DAGs in Figures~\ref{f5}(b)--\ref{f5}(f), the entire vector
$\mathbf{C}$ is not necessary to identify the effect of $\mathbf{A}$
on $Y_{i}$, though it would in general be necessary in order to jointly
identify the effects of $\mathbf{A}$ on $Y_{i}$ and on $Y_{j}$
(i.e., the effect of $\mathbf{A}$ on $\mathbf{Y}$). This
is because in order to identify the effect of $\mathbf{A}$ on
$\mathbf{Y}$ we require conditional exchangeability to hold for
all subjects. In these settings, if $\mathbf{C}$ is not fully
observed but certain components or functions of $\mathbf{C}$
required to identify the effect of $\mathbf{A}$ on $Y_{i}$ are,
then we can proceed by considering $Y_{i}$ to be the only outcome
in each block. We can still consider the full vector of treatments
$\mathbf{A}$, and therefore identify unit-level, spillover, total
and overall effects of $\mathbf{A}$ on~$Y_{i}$.

%s4 #&#
\section{Three Distinct Types of Interference}\label{s4}

By understanding the causal mechanisms underlying interference, we
can more precisely target effects of interest.
There are three distinct
causal pathways by which one individual's treatment may affect another's
outcome. All fall under the rubric of interference. The distinction
among them has generally not been made, but they differ in causal
structure, effects of interest and requirements for identification
of effects. Often more than one type of interference will be present
simultaneously.

The first pathway by which interference may operate is a direct causal
effect of one individual's treatment on another individual's outcome,
unmediated with respect to the first individual's outcome. We call
this \emph{direct interference}. As an example, suppose that the outcome
is obesity and the treatment is dietary counseling from a nutritionist.
An individual who receives treatment can in turn ``treat'' his associates
by imparting to them the information gained from the nutritionist;
therefore, if individual $i$ receives treatment and individual $j$
does not, individual $j$ may be nevertheless be exposed to the treatment
of individual $i$ and his or her outcome will be affected accordingly.

A second pathway by which one individual's treatment may affect another
individual's outcome is via the first individual's outcome. For example,
if the outcome is an infectious disease and the treatment is a prophylactic
measure designed to prevent disease, then the treatment of individual
$i$ may affect the outcome of individual $j$ by preventing individual
$i$ from contracting the disease and thereby from passing it on.
We call this type of interference \emph{interference by contagion}.
It is differentiated from direct interference by the fact that it
does not represent a direct causal pathway from the exposed individual
to another individual's outcome, but rather a pathway mediated by
the outcome of the exposed individual.

The third pathway for interference is \emph{allocational interference}.
Treatment in this setting allocates individuals to groups; through
interactions within a group individuals' characteristics may affect
one another. An example that often arises in the social science literature
is the allocation of children to schools or of children to classrooms
within schools (\citeauthor{angrist2004does}, \citeyear
{angrist2004does}; Graham, Imbens and Ridder, \citeyear
{graham2010measuring};
\citeauthor{hong2008causal}, \citeyear{hong2008causal}).
The performance and behavior of student $i$ may affect the performance
and behavior of student $j$ in the same class, for example, by distracting
or motivating student $j$ or by occupying the teacher's attention.
Another example that can be seen as allocational interference is the
effect of college enrollment on wage differences for college- versus
high-school-educated workers, where the wage difference depends on
the proportion of workers in each education category
(\citeauthor{heckman1998explaining}, \citeyear{heckman1998explaining}).

%s4.1 #&#
\subsection{Direct Interference}\label{s4.1}

Direct interference is present when there is a causal pathway from
one individual's treatment to another individual's outcome, not mediated
by the first individual's outcome. Interference can be direct with
respect to a particular outcome but not another. Consider two individuals
living in the same household, each randomized to an intervention designed
to prevent high cholesterol. Suppose the intervention consists of
cooking classes, nutritional counseling and coupons that can be redeemed
for fresh produce, and consider a household in which one individual
is treated and one untreated. The treated individual could bring fresh
produce into the household, prepare healthy meals, and talk about
the nutritionist's counsel, thereby exposing the other individual
to a healthier diet. If the outcome of interest is a measure of blood
cholesterol level, then this is an example direct interference: the
untreated individual is exposed to the treated individual's diet and
that exposure reduces the untreated individual's cholesterol. On the
other hand, if the outcome is a measure of healthy diet and behavior,
then the same story depicts contagion rather than direct interference:
the treated individual adopts a healthier diet which results in the
untreated individual also adopting a healthier diet. Diet may spread
by contagion; cholesterol presumably would not.

In many settings, direct interference and contagion will be present
simultaneously for the same outcome. For example, suppose that in
the story above the outcome were weight change. Then it is possible
that the treated individual's family member could lose weight both
because of exposure to healthier foods (direct interference) and because
he was motivated by the weight loss of his relative (contagion).

Direct interference has the simplest causal structure of the three
types of interference. In addition to a direct causal path from $A_{i}$
to $Y_{i}$, there is also a direct path from $A_{i}$ to $Y_{j}$
for all pairs $(i,j)$ such that subjects $i$ and $j$ are in the
same block. Direct interference in a block of size two is depicted
in the DAGs in Figures~\ref{f4} and \ref{f5}, with the exception of
Figure~\ref{f5}(g). Because
there is only a single path from $A_{i}$ to $Y_{j}$ for any pair
$i,j$, differences between counterfactuals of the form $Y_{i}(\mathbf{a})$
capture all of the causal effects of $\mathbf{A}$ on $Y_{i}$
and, therefore, effects like the total, unit-level, spillover and overall
effects described in Section~\ref{s2.1} summarize the causal effects
of $\mathbf{A}$
on~$Y_{i}$.

%s4.2 #&#
\subsection{Interference by Contagion}\label{s4.2}

\begin{figure}[t]

\includegraphics{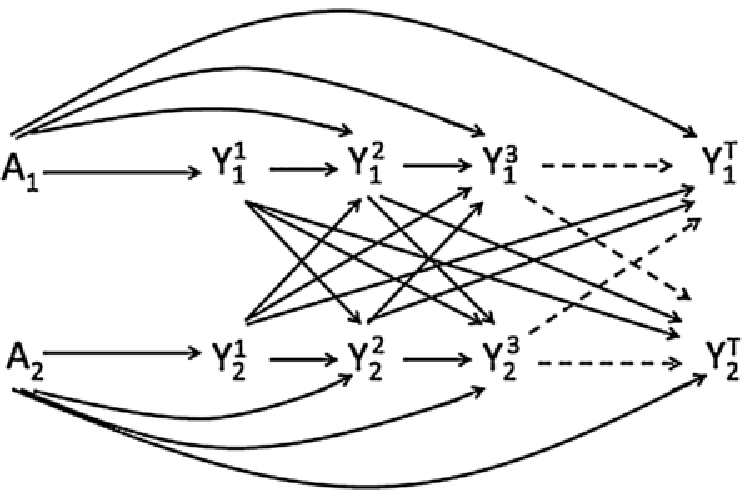}

\caption{}\label{f6}
\end{figure}

Interference by contagion often has a complex causal structure, because
it can involve feedback among different individuals' outcomes over
time. The causal structure of the effect of $A_{i}$ on $Y_{i}$ is
straightforward: $A_{i}$ has a direct protective effect on $Y_{i}$,
represented by a direct arrow from $A_{i}$ to $Y_{i}$ on the DAG.
The effect of $A_{i}$ on $Y_{j}$ is considerably more complex. It
is tempting to represent the effect of $A_{i}$ on $Y_{j}$ as a mediated
effect through $Y_{i}$, but this cannot be correct, as $Y_{i}$ and
$Y_{j}$ are contemporaneous and, therefore, one cannot cause the other.
The effect of $A_{i}$ on $Y_{j}$ is mediated through the evolution
of the outcome of individual $i$; this complicated structure is depicted
in the DAG in Figure~\ref{f6}, where $Y_{i}^{t}$ represents the outcome
of individual $i$ at time $t$, $T$ is the time of the end of follow-up,
and the dashed arrows represent times $4$ through $T-1$, which do
not fit on the DAG (but which we assume were observed). The unit of
time required to capture the causal structure depends on the nature
of transmission of the outcome; it should be the case that the probability
of one individual's outcome affecting another's is unaltered by differences
in timing on a scale smaller than the units used.

%f6 #&#
%

In order to further explicate the structure, we consider the case of
an infectious disease like the flu. Infectious diseases are paradigmatic
examples of contagion. \citet{halloran1995causal}, \citet
{hudgens2008toward}
and \citet{vanderweele2011bounding} have written about identification
and estimation of overall, unit-level, spillover and total effects
for vaccinations against infectious diseases, and we follow up with
this literature in Section~\ref{s4.2.1}. Although we illustrate the principles
of interference by contagion through the lens of infectious diseases,
this type of interference can occur in many and diverse settings:
an educational intervention assigned to one student could affect that
student's performance, which in turn might affect the performance
of her classmates; a get-out-the-vote mailing could motivate its recipients
to decide to vote, and communicating that decision to friends could
change the friends' voting behavior. The principles that we discuss
below apply to any of these settings.

Suppose that the flu vaccine has a protective effect against the flu
by preventing or shortening the duration of episodes of the flu for
some individuals. Let $A$ be an indicator of getting the flu vaccine
before the start of a six-month long flu season, and let $Y$ be the
total number of days spent infectious with the flu over the course
of the season. In the DAG in Figure~\ref{f6}, $Y_{i}^{t}$ represents the
flu status of individual $i$ at time $t$ (measured in days), $T\equiv180$
is the day of the end of flu season, and the dashed arrows represent
days $4$ through $T-1$, which do not fit on the DAG (but which we
assume were observed). Let $Y_{i}^{t}$ be the total number of days
spent infectious up to and including day $t$. (Note that, when $Y_{i}^{t}$
is observed for all $t$, this is equivalent to coding it as an indicator
of individual $i$ being infectious at time $t$.) In choosing days
as the unit of time, we are making the assumption that the probability
of one individual infecting another is not affected by a difference
of a fraction of a day in flu duration.

%f7 #&#
%
\begin{figure}[b]

\includegraphics{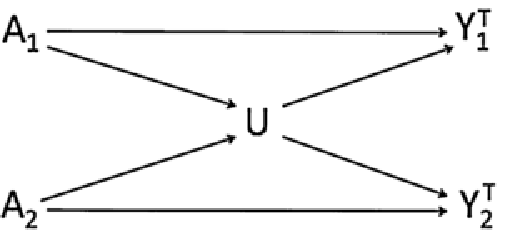}

\caption{}\label{f7}
\end{figure}

We will rarely have fine-grained information on the evolution of the
outcome over time. In the rest of this section, we describe how to
draw the appropriate causal DAGs and how to identify causal effects
in such cases. Drawing a causal DAG using only a subset of relevant
variables has been extensively studied in the graphical models literature
and involves an operation known as projection (\citeauthor
{pearl1995theory}, \citeyear{pearl1995theory};
\citeauthor{tian2002testable}, \citeyear{tian2002testable};
\citeauthor{verma1993graphical}, \citeyear{verma1993graphical}).
Projection algorithms are somewhat technical; below we provide an
intuitive discussion of the construction of causal DAGs when not all
variables relevant to contagion are observed.

\begin{figure}%[t]

\includegraphics{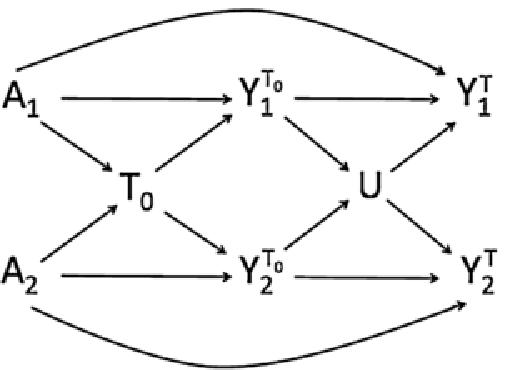}

\caption{}\label{f8}
\end{figure}

If we only observe the outcome (cumulative days of the flu) at the
end of the season, then, as in the DAG in Figure~\ref{f7}, we replace the
collection of all unobserved variables (i.e., $Y_{i}^{t}$ for $t<T$)
with $U$. Without additional assumptions, we cannot replace the two
diagonal pathways through $U$ with direct arrows from $A_{1}$ to
$Y_{2}^{T}$ and from $A_{2}$ to $Y_{1}^{T}$; that would imply that
$ Y_{1}^{T}\amalg Y_{2}^{T} |A_{1},A_{2}$, which is shown
to be false by the DAG in Figure~\ref{f6}. If we know who gets the flu first,
then the DAG in Figure~\ref{f8} represents the causal relations among the
observed variables, where $T_{0}$ is the time of the first case of
the flu. The unmeasured variable $U$ cannot be omitted because
$Y_{1}^{T}$
is not independent of $Y_{2}^{T}$ conditional on
$ \{ Y_{1}^{T_{0}},Y_{2}^{T_{0}} \} $;
$Y_{1}^{T}$ depends on the number and timing of individual 2's illnesses
between time $T_{0}$ and time $T$. It might seem as though $Y_{1}^{T_{0}}$
should be independent of $Y_{2}^{T_{0}}$ conditional on $ \{
A_{1},A_{2} \} $
in this scenario, because there can be no contagion before the first
case of the flu. But this is not the case: $Y_{i}^{T_{0}}$ can be
thought of as an indicator that individual $i$ gets the flu before
or at the same time as individual $j$, and this is dependent on individual
$j$ remaining healthy through time $T_{0}-1$. On the other hand,
conditioning on the time of the first case of the flu renders $Y_{1}^{T_{0}}$
and $Y_{2}^{T_{0}}$ independent, because conditioning on $T_{0}$
is tantamount to conditioning on both individuals remaining healthy
until the time of the first case, that is, on their entire flu histories
up to time $T_{0}$.

%f8 #&#
%

\begin{figure}%[t]

\includegraphics{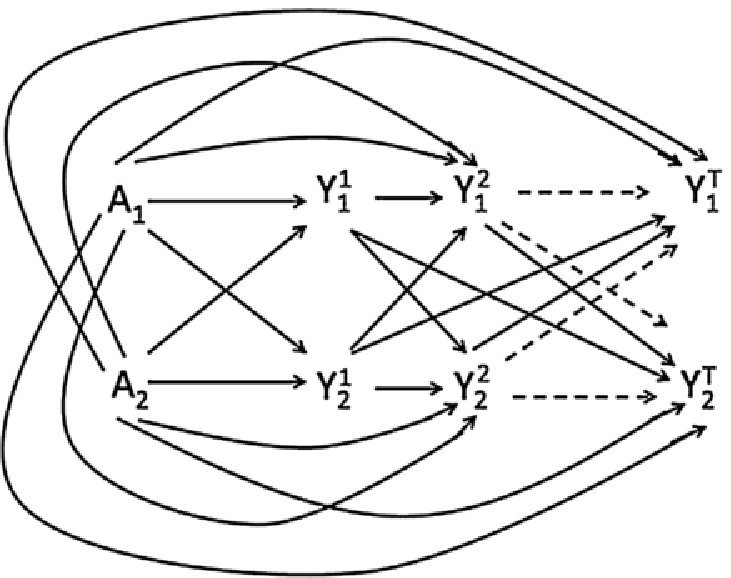}

\caption{}\label{f9}
\end{figure}

Suppose information on the number but not duration of cases of the
flu is available. Then we could define $Y^{t}$ to be the number of
distinct cases initiated by time $t$. However, this outcome fails
to capture all of the relevant information about an individual's flu
status, because a case of the flu that lasts ten days may be more
contagious than one that lasts five days. Therefore, there may be
an effect of $A_{i}$ on $Y_{j}^{t}$ that is not mediated by $Y_{j}^{s}$,
$s<t$, being instead mediated by the duration of individual $i$'s
flu incidents. This is represented on the DAG in Figure~\ref{f9} by an arrow
from $A_{i}$ to $Y_{j}^{t}$. These arrows encode the fact that $Y_{j}^{t}$
is dependent on $Y_{i}^{t}$ even conditional on $ \{
Y_{i}^{s},Y_{j}^{s} \} $
for all $s<t$. Similarly, if the outcome on the DAG in Figure~\ref{f8} were
the total number of flu episodes by the end of the season instead
of the cumulative days of flu we would add arrows from $A_{i}$ to
$Y_{j}^{T_{0}}$ and to $Y_{j}^{T}$.

%f9 #&#
%

%f10 #&#
%
\begin{figure}[b]

\includegraphics{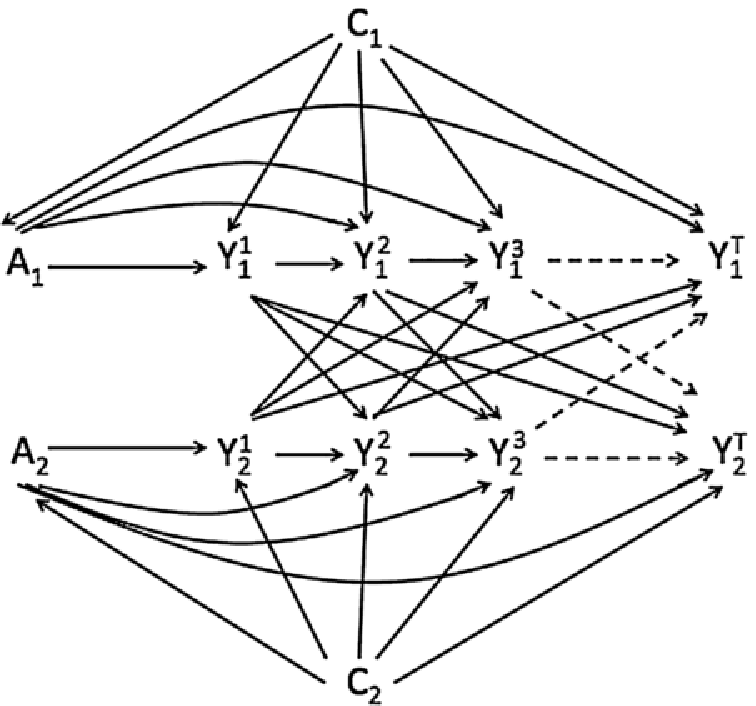}

\caption{}\label{f10}
\end{figure}

If there are common causes $C$ of treatment and outcome for each
individual, as in the DAG in Figure~\ref{f10}, then exchangeability
for the
effect of $\mathbf{A}$ on $Y_{i}^{T}$ will hold conditional
on $C_{i}$. If there are common causes of treatments for different
individuals, or of treatments and outcomes across individuals, then
exchangeability requires conditioning on them. The same conclusions
about exchangeability hold if we observe the outcome only at select
time points.

The overall, unit-level, spillover and total effects defined in
Section~\ref{s2.1} do not distinguish among the multiple distinct
causal pathways
from $A_{i}$ to $Y_{j}^{T}$. We discuss estimation of path-specific
effects below.

%s4.2.1 #&#
\subsubsection{Contagion and infectiousness}\label{s4.2.1}

Recently, \citet{vanderweele2011components} described the
decomposition
of the spillover effect, that is the effect of $A_{i}$ on $Y_{j}$,
into a ``contagion effect'' and an ``infectiousness effect.''
The contagion effect is the protective effect that vaccinating one
individual has on another's disease status by preventing the vaccinated
individual from getting the disease and thereby from transmitting
it, similar to the effect we discussed above. In order to illustrate
the infectiousness effect, consider the following refinement to our
example. Suppose that a vaccine exists for the flu that prevents the
disease for some individuals and also makes some cases of the disease
among vaccinated individuals less likely to be transmitted. Let $A$
be an indicator of vaccination before the start of flu season and
let $Y^{t}$ be the total number of episodes of flu up to and including
day $t$. Then $A_{i}$ may have an effect on $Y_{j}^{t}$ even if
it has no effect on $Y_{i}^{t}$, that is, even if individual $i$ would
get the flu whether vaccinated or not, by preventing individual $i$
from transmitting the flu to individual $j$. This infectiousness
effect represents a pathway that is distinct from the contagion effect
because it does not operate through the infection status of the vaccinated
individual. The infectiousness effect has the structure of a direct
effect by which individual $i$'s vaccination confers improved protection
against individual $i$'s flu on individual $j$; it represents a
type of direct interference. This is similar to the example above
in which the duration of flu episodes was unobserved: infectiousness,
like flu duration, is a property of the infected individual's disease
state, but if it is not captured by the outcome measure then it has
the structure of a direct effect of $A_{i}$ on $Y_{j}$. The contagion
and infectiousness effects are not identifiable without strong assumptions
or carefully conceived data collection. When they are identifiable
their sum (or product if they are defined on the multiplicative scale)
is equal to the total spillover effect of $A_{i}$ on $Y_{j}$.

%f11 #&#
%
\begin{figure}

\includegraphics{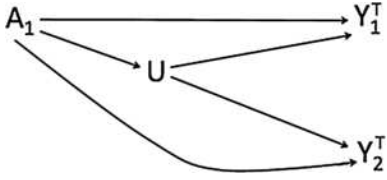}

\caption{This DAG corresponds to a household of size two in which
individual 2 is always unvaccinated and disease status is assessed
at the end of follow-up. $U$ represents unmeasured variables.}\label{f11}
\end{figure}

%f12 #&#
%
\begin{figure}[b]

\includegraphics{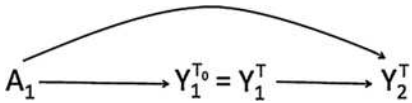}

\caption{This DAG corresponds to the same setting as Figure~\protect
\ref{f11},
but under the assumptions that individual 2 can only be infected by
individual 1 and that only one event is possible per subject during
follow-up.}\label{f12}
\end{figure}

\citet{vanderweele2011components} defined the contagion and infectiousness
effects as the natural indirect and direct effects, respectively,
of $A_{1}$ on $Y_{2}^{T}$ with $Y_{1}^{T_{0}}$ as the mediator,
where $T_{0}$ is the day of the first flu infection and $T$ is the
day of the end of follow-up, for example, the last day of the flu season.
That is, the contagion effect is defined as $E [Y_{2}^{T}
(0,Y_{1}^{T_{0}}(1) ) ]-E [Y_{2}^{T} (0,Y_{1}^{T_{0}}(0) ) ]$
and infectiousness by $E [Y_{2}^{T} (1,Y_{1}^{T_{0}}(1) ) ]-E
[Y_{2}^{T} (0,Y_{1}^{T_{0}}(1) ) ]$.
For simplicity and consistency with the existing literature we adopt
the setting used in \citet{vanderweele2011bounding},
\citet{vanderweele2011components}
and \citet{halloran2012}: each block is a group of size two who share
a household, and in each pair individual 1 is randomized to vaccine
while individual 2 is always unvaccinated. If, as those authors assumed,
disease status is observed only at the end of follow-up, Figure~\ref{f11}
depicts the operative DAG. Although there is an unmeasured variable
on the path from $A_{1}$ to $Y_{2}^{T}$, it is not a confounder
and we can identify the effect of $A_{1}$ on $Y_{2}^{T}$. However,
we cannot identify the component contagion and infectiousness effects
without observing the mediator $Y_{1}^{T_{0}}$. In order to circumvent
the problem of the unobserved mediator, \citet{vanderweele2011components}
assumed that each individual can be infected only once and that individual
2 can only be infected by individual 1, as would be the case if individual
2 were homebound. These assumptions dramatically simplify the causal
structure by ensuring that individual 1 is infected first and that
there is no feedback between the time of first infection and the end
of follow up. Then $Y_{1}^{T}$ must be equal to $Y_{1}^{T_{0}}$,
and thus $Y_{1}^{T_{0}}$ is observed. These assumptions are encoded
by the DAG in Figure~\ref{f12}. Now the contagion and infectiousness effects
can be identified as the natural indirect and direct effects of $A_{1}$
on $Y_{2}^{T}$ mediated by $Y_{1}^{T}=Y_{1}^{T_{0}}$, as long as
assumptions (\ref{med no int 1}) through (\ref{med no int 3}) are
met. Three of these assumptions correspond to the absence of any unmeasured
confounding of the relationships between $A_{1}$ and $Y_{1}^{T}$,
between $Y_{1}^{T}$ and $Y_{2}^{T}$ conditional on $A_{1}$, and
between $A_{1}$ and $Y_{2}^{T}$, respectively. These assumptions
can be made conditional on measured covariates $\mathbf{C}$.
Assumption (\ref{med no int 4}) is the recanting witness criterion.

%f13 #&#
%
\begin{figure}[b]

\includegraphics{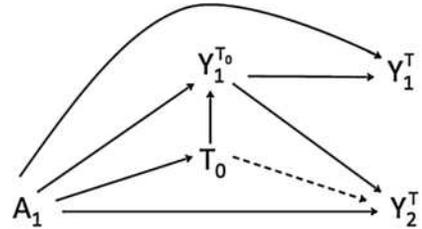}

\caption{This DAG corresponds to the same setting as Figure~\protect
\ref{f12},
but without the assumption that individual 2 can only be infected
by individual 1. The dashed arrow is present when $T$ is defined
as the end of the flu season; it is absent when $T$ is defined as
$T_{0}+s$.}\label{f13}
\end{figure}

The simplifying assumption that the outcome can occur at most once
may be reasonable for many infectious diseases. More limiting is the
assumption that individual 2 can only be infected by individual 1.
Here, we will describe settings in which it may be possible to relax
this assumption. Even if we observe the time and identity of the first
case of the flu in addition to the outcome at the end of follow-up,
relaxing this assumption makes identification of the contagion and
infectiousness effects impossible. In the DAG in Figure~\ref{f13},
$Y_{1}^{T_{0}}$
is an indicator of whether individual 1 was infected first. Although
it is not straightforward to imagine intervening on $T_{0}$, the
time of the first infection, we include this variable in the DAG in
order to ensure that the DAG encodes the true conditional independence
statements. The presence of the arrow from $T_{0}$ to $Y_{2}^{T}$
makes $T_{0}$ a recanting witness for the contagion and infectiousness
effects: it is a confounder of the mediator--outcome relation that
is caused by treatment. This arrow is necessitated by the fact that
$T_{0}$ predicts $Y_{2}^{T}$ even conditional on $A_{1}$ and $
Y_{1}^{T_{0}}$.
To see this, imagine two different pairs in which individual 1 is
vaccinated and gets sick first ($A_{1}=Y_{1}^{T_{0}}=1$). Suppose
that in one pair, the vaccinated individual gets sick at the very
end of the follow-up period ($T_{0}=T-1$). Then the probability that
the second individual gets the flu after time $T_{0}$ but before
time $T$ is very small. Suppose that in the other pair the vaccinated
individual gets sick on the first day of the flu season ($T_{0}=1$).
The probability that we observe the second individual in this pair
to get sick before the end of follow-up is much higher.

One possible solution to the recanting witness problem is to let $T_{0}$
determine a new artificial end of follow-up, so that the amount of
time between $T_{0}$ and $T$ is constant over different values of
$T_{0}$. In particular, if we know that an infected individual is
infectious for up to $s$ days after becoming symptomatic, then we
can let $T=T_{0}+s$ and collect data on $Y_{1}^{T_{0}}$, $Y_{1}^{T_{0}+s}$
and $Y_{2}^{T_{0}+s}$. If neither individual in the pair is observed
to get the flu then $T_{0}=T=$ \emph{the last day of the flu season}.
Setting the artificial end of follow-up to lag behind the time of
first infection by $s$ days ensures that we will observe
$Y_{1}^{T_{0}}=Y_{2}^{T_{0}+s}=1$
for any pair in which individual 2 catches the flu from individual
1. We throw away data on pairs for which the first infection occurs
fewer than $s$ days before the end of the flu season, but if $s$
is small then it may be reasonable to assume that any resulting bias
is negligible.

One further assumption is required in order for
$ Y_{2}^{T_{0}+s}\amalg T_{0} |Y_{1}^{T_{0}},A_{1}$,
which is the conditional independence assumption that licenses the
omission of an arrow from $T_{0}$ to $Y_{2}^{T_{0}+s}$. Suppose
that cumulative exposure to the flu virus makes people, on average,
less susceptible to infection as the flu season progresses due to
acquired immunity. Then, for pairs in which individual 1 is vaccinated
and gets sick first, individual 2 is less likely to catch individual
1's flu later in the season as compared to earlier (for larger values
$T_{0}$ compared to smaller values). This violates
$ Y_{2}^{T_{0}+s}\amalg T_{0} |Y_{1}^{T_{0}},A_{1}$.
If we assume that the probability of individual 2 catching the flu
if exposed on day $t$ is constant in $t$, then $Y_{2}^{T_{0}+s}$
is independent of $T_{0}$ conditional on $A_{1}$ and $
Y_{1}^{T_{0}}$.
Therefore, $T_{0}$ is not a recanting witness (it is still caused
by treatment but is no longer a confounder of the mediator--outcome
relation). Assuming that exchangeability assumptions (\ref{med no int 1}),
(\ref{med no int 2}) and (\ref{med no int 3}) hold (see
\citeauthor{vanderweele2011components}, \citeyear{vanderweele2011components}
for discussion of their plausibility in this context), the contagion
and infectiousness effects of $A_{1}$ on $Y_{2}^{T_{0}+s}$ are identifiable.
The contagion effect is given by $E [Y_{2}^{T_{0}+s}
(0,Y_{1}^{T_{0}}(1) ) ]-E [Y_{2}^{T_{0}+s} (0,Y_{1}^{T_{0}}(0) ) ]$
and infectiousness by $E [Y_{2}^{T_{0}+s} (1,Y_{1}^{T_{0}}(1) ) ]
-E [Y_{2}^{T_{0}+s} (0,\break  Y_{1}^{T_{0}}(1) ) ]$.
The spillover effect of $A_{1}$ on $Y_{2}^{T_{0}+s}$ on the additive
scale is the sum of the contagion and infectiousness effects.

If both individuals are randomized to vaccination, then $A_{2}$ is
a confounder of the relationship between $Y_{1}^{T_{0}}$ and $Y_{2}^{T_{0}+s}$.
Assuming $A_{2}$ is observed, this does not pose a problem for identification
of the contagion and infectiousness effects. Generalizations of the
contagion and infectiousness effects to blocks of size greater than
two are also possible (\citeauthor{vanderweele2011components},
\citeyear{vanderweele2011components}).

An alternative definition of an infectiousness effect, proposed by
\citet{vanderweele2011components}, is the controlled direct effect
of $A_{1}$ on $Y_{2}^{T}$, holding $Y_{1}^{T_{0}}$ fixed at $1$.
Identification of this effect does not require the recanting witness
criterion to hold and, therefore, it is identifiable when the end-of-follow
up is fixed and does not depend on $T_{0}$. A disadvantage of this
controlled direct infectiousness effect is that it does not admit
a decomposition of the indirect effect of $A_{1}$ on $Y_{2}^{T}$.
That is, if we subtract the controlled direct infectiousness effect
from the total effect of $A_{1}$ on $Y_{2}^{T}$, the remainder
cannot be interpreted as a contagion effect.

%s4.3 #&#
\subsection{Allocational Interference}\label{s4.3}

In allocational interference, an individual is allocated to a group
and his outcome is affected by which individuals are allocated to
the same group. In many real settings, group allocation is not random,
rather individuals select their own group or are assigned based on
previously observed characteristics. One example of random group allocation
is the assignment of college freshman to dorms or dorm rooms
(\citeauthor{carrell2009does}, \citeyear{carrell2009does};
\citeauthor{sacerdote2000peer}, \citeyear{sacerdote2000peer}).
We differentiate between allocational interference and other scenarios
in which individuals already in groups are assigned to receive individual
or group-level treatments. In the former setting, an individual's outcome
depends on the specific composition of his group, while in the latter
it depends on treatment assignments for his group but not necessarily
on group composition. Of course, these two phenomena often occur in
tandem, as they would if children were assigned to classrooms which
were then assigned different educational interventions.

Similarly, contagion is often present in conjunction with allocational
interference. For example, in the allocation of children to classrooms
there is likely to be feedback among children who are in the same
classroom in terms of achievement, attitude and tendency to act out.
Therefore, any measure of behavior or achievement that can evolve over
time would likely be subject to contagion. An outcome like end-of-the-year
test scores, on the other hand, would evince allocational interference
without contagion as one student's test score cannot directly affect
another's (though we would likely still envision contagion with respect
to knowledge or learning).

Allocational interference is perhaps the most complicated of the three
types of interference. We first describe how to represent basic allocational
interference on a DAG. Then we introduce a toy example and use it
to illustrate some additional DAG structures and to briefly discuss
causal effects that may be of interest in the presence of allocational
interference. This discussion is far from exhaustive, but we hope
that this section will serve as a guide for how to think about allocational
interference.

Allocational interference assigns subjects to groups within each block.
Recall that blocks of individuals are independent from one another,
and that interference is possible within but not between blocks. This
is not the case for groups; interference will generally be present
across groups within the same block. In the school example, blocks
might be different schools and groups might be classrooms within each
school. Suppose that within each block there are $L$ possible groups
to which an individual can be allocated. Then treatment is a categorical
variable, $A$, which for each subject takes on the value $l\in
(1,\ldots,L )$
of the group to which that individual was assigned. As in Section~\ref{s2},
we let $k$ index $N$ blocks with $m$ individuals in each block.
Let $\mathbf{A}_{k}$ be the vector of group assignments in block
$k$.

%f14 #&#
%
\begin{figure}[b]

\includegraphics{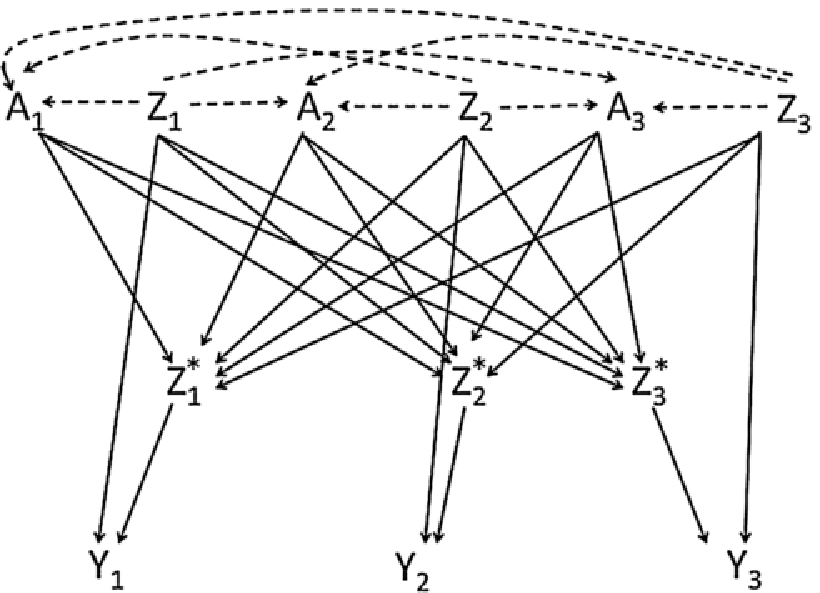}

\caption{}\label{f14}
\end{figure}

Figure~\ref{f14} provides a DAG for a scenario with allocational interference
and $m=3$. Within each block, we allocate the subjects into two distinct
groups and each individual's outcome may be affected by who is in
each group. The DAG depicts a single block and we therefore suppress
the subscript $k$. Let $Y_{i}$ be the outcome for individual $i$
and $Z_{i}$ be a vector of all baseline characteristics that affect
the outcome of individual $i$ or the outcomes of the other individuals
with whom he comes into contact. Define $Z_{i}^{*}$ to be an $(m-1)$-dimensional
array with $j$th element equal to $Z_{j}\times I(A_{j}=A_{i})$,
$j\neq i$, that is, the baseline covariates of subject $j$ multiplied
by the indicator that individual $j$ is assigned to the same group
as individual $i$. In addition to the individual level causal arrows
from $Z_{i}$ and $Z_{i}^{*}$ into $Y_{i}$, we require arrows from
$Z_{i}$ to $Z_{j}^{*}$ and from $A_{i}$ to $Z_{j}^{*}$ for all
pairs $(i,j)$, $i\neq j$. This is because $Z_{j}^{*}$ is by definition
a function of $\mathbf{A}$ and of $Z_{i}$ for all $i\neq j$.
Randomized allocation corresponds to the absence of the dashed arrows
into $\mathbf{A}$. Otherwise, an individual's baseline covariates
$Z_{i}$ may affect his group assignment $A_{i}$. For $Z_{i}$ to
affect $A_{i}$ and not $A_{j}$ would entail an allocation rule that
ignores balance across groups. In some settings, the vector of baseline
covariates $\mathbf{Z}$ (or a function of $\mathbf{Z}$,
e.g., its mean) would affect the allocation rule. This is represented
by the presence of the dashed arrows into~$\mathbf{A}$.

We now describe a toy example in which allocational interference operates
and informs causal effects of interest. Suppose that runners enter
a 5000 meter race, but the track is not wide enough for all of the
competitors to race simultaneously. A race represents a single interference
block; in order to perform statistical inference on the effects discussed
below we would likely need to observe several independent races, indexed
by $k$. The runners in each race are divided into smaller groups
to race in successive heats. Number the subjects according to some
composite measure of their recent performance, so that runner 1 is
the fastest based on the composite measure and runner $m$ is the
slowest. Let $Y_{ki}$ be the time in which runner $i$ in block $k$
finishes the race and $Z_{ki}$ be a vector of all relevant baseline
characteristics. A runner's speed will affect his own outcome. Moreover,
his speed, confidence and sportsmanship may have an impact on the
outcomes of the runners with whom he is grouped. These characteristics
should all be included in $Z_{ki}$. The runners are divided into
three heats, so $A_{ki}\in\{ 1,2,3 \} $. For simplicity,
we assume that $m$ is divisible by $3$ and each heat has $m/3$
runners, though heats of different sizes are possible. Consider the
following two allocations: In allocation $\mathbf{a}$, runners
1 through $m/3$ are assigned to the first heat, runners $(m/3)+1$
through $2m/3$ to the second heat and runners $(2m/3)+1$ through
$m$ to the third heat. In allocation $\mathbf{a}'$, on the other
hand, the first heat is comprised of runners $1, 4, 7, \ldots, m-2$;
the second of runners $2, 5, 8, \ldots, m-1$, and the third of the
remaining runners. Allocation $\mathbf{a}'$ results in a more
balanced distribution of baseline speed across heats.

Are runners, on average, likely to run faster under one of these two
allocations? This is a question about the overall causal effect
$E[Y(\mathbf{a})]-E[Y(\mathbf{a}')]$.
If the number of groups is the same under both allocations, as in
our example, then the direct and indirect effects of allocations
$\mathbf{a}$
and $\mathbf{a}'$ may also be of interest (see Section~\ref{s2.1}).
The expected unit-level effect $\mathit{UE}_{1}(\mathbf{a};3,1)\equiv
E[Y_{1}(\mathbf{a}_{-1},3)]-E[Y_{1}(\mathbf{a}_{-1},1)]$
is the expected effect on runner 1 of racing in the fastest versus
the slowest heat in allocation $\mathbf{a}$. The expected spillover
effect $\mathit{SE}_{1}(\mathbf{a},\mathbf{a}';1)\equiv
E[Y_{1}(\mathbf{a}_{-1},1)]-
  E[Y_{1}(\mathbf{a}_{-1}',1)]$
is the expected effect on runner 1 of running in the first heat when
that heat is comprised of the fastest runners versus running in the
first heat when that heat is comprised of runners with a mix of speeds.
(In both cases, the expectations are with respect to multiple independent
races.) This might matter if running in the first heat was advantageous
because the crowd was more enthusiastic earlier on, a point to which
we return below.

As always, in order to identify expectations of counterfactuals of
the form $Y_{ki}(\mathbf{a}_{k})$ we require conditional exchangeability
for the effect of $\mathbf{A}_{k}$ on $Y_{ki}$. If $\mathbf{A}_{k}$
and $Z_{ki}$ share any common causes, as they would if, for example,
heats were assigned based on the identity of the runners' coaches,
then those common causes must be included in the conditioning set.
If $A_{ki}$ depends on any component of $Z_{ki}$ then that component
must be included in the conditioning set in order to achieve exchangeability.
Similarly, if $A_{ki}$ depends on a component of $\mathbf{Z}_{k}$,
that is, there are arrows from $Z_{ki}$ and from $Z_{kj}$, $j\neq i$,
into $A_{ki}$, then that component of $\mathbf{Z}_{k}$ must
be included in the conditioning set. We note that conditioning on
a component of $\mathbf{Z}_{k}$ may block part of the effect
of $\mathbf{A}_{k}$ on $Y_{ki}$; because $Z_{ki}^{*}$ is a
deterministic function of $\mathbf{Z}_{k}$ conditioning on the
latter is effectively conditioning on the former. $Z_{ki}^{*}$ lies
on the causal pathway from $\mathbf{A}_{k}$ to $Y_{ki}$ and,
therefore, conditioning on it blocks part of the causal effect of interest.

%f15 #&#
%
\begin{figure}[b]

\includegraphics{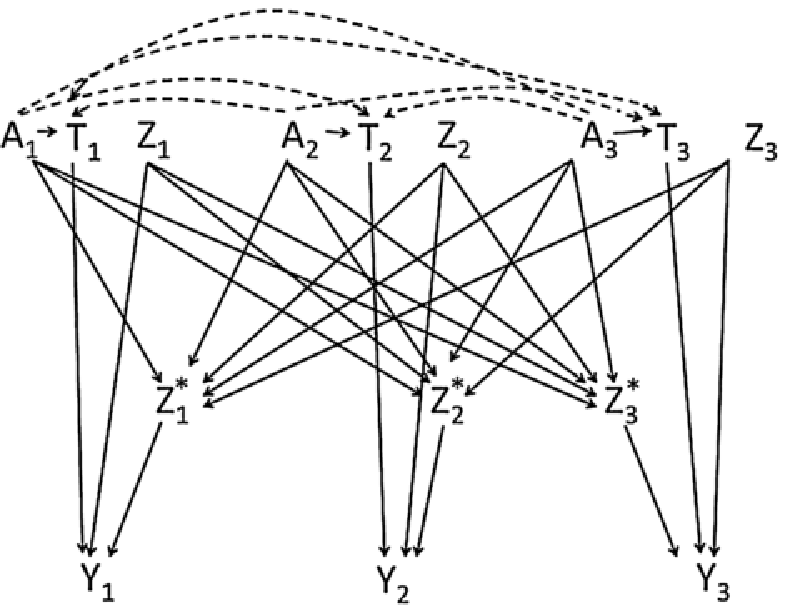}

\caption{}\label{f15}
\end{figure}

Let $T_{kl_{i}}$ be a group-level property of group $l_{i}$ in block
$k$, where $l_{i}$ indexes the group to which individual $i$ is
assigned, and define $T_{ki}\equiv T_{kl_{i}}$ to be the group-level
property to which individual $i$ is exposed. If the order in which
the heats are run makes a difference, because the weather changes
throughout the day or because runners are tired later in the day,
then $T_{ki}$ could be the time at which subject $i$'s heat is scheduled
to race. This is an example of a preallocation group-level covariate
that allows the groups to be distinguished from one another without
reference to their composition. Other examples of this kind of group-level
property are the teacher in charge of each classroom, the curricula
to which classrooms are assigned, or different locations in which
heats are assigned to run.

\begin{figure}[t]

\includegraphics{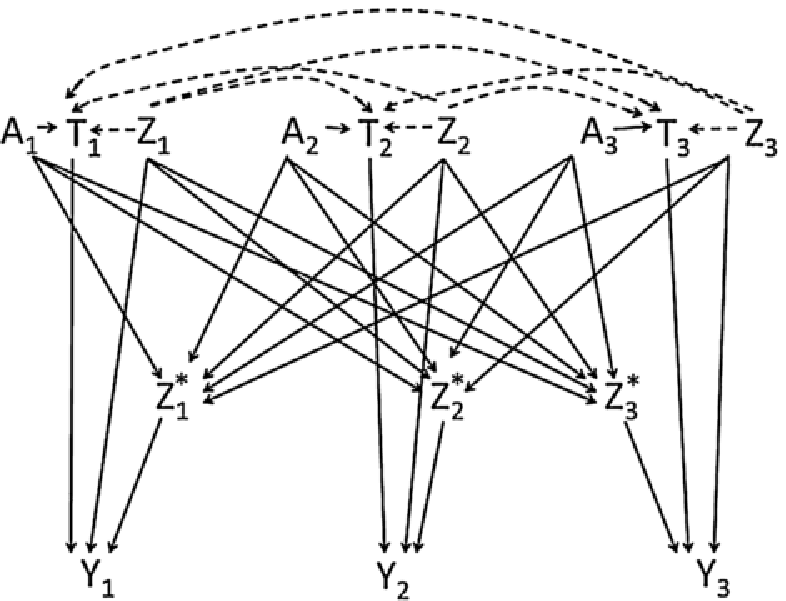}

\caption{}\label{f16}
\end{figure}

For the purposes of our race example, let $T_{ki}$ be a measure of
the crowd enthusiasm when runner $i$ runs. The arrows from $A_{i}$
into $T_{i}$ on the DAG in Figure~\ref{f15} are necessitated by the way
we have defined $T_{ki}$, namely as a collection of properties of
the group to which individual $i$ is assigned. Properties that depend
on group composition but are not captured by $Z_{ki}^{*}$, such as
the number of runners in the heat, can also affect $Y_{ki}$. Unlike
the time at which each heat runs, these properties arise after group
allocation and, therefore, do not distinguish the groups from one another
a priori. If $T_{ki}$ is itself affected by group composition, because
the size of the crowd is determined by who is in each heat, then we
would also require arrows $A_{j}\rightarrow T_{i}$; these are the
dashed arrows in Figure~\ref{f15}. Suppose that crowd enthusiasm is
determined
by the proportion of runners in the heat who are in the fastest
quartile
of all of the runners in the race, based on the baseline composite
measure. Then $T_{ki}$ is affected by $\mathbf{Z}_{k}$ (which
includes a measure of each runner's previous performance), and we
require arrows from each $Z_{j}$ into $T_{i}$, as on the DAG in
Figure~\ref{f16}. If data on $\mathbf{T}_{k}$ is not collected, or if
it is not known whether any group-level properties affect $Y_{ki}$,
then we would add arrows from each $A_{i}$ into each $Y_{j}$ on
the DAG in Figure~\ref{f15}, to represent the residual effect
of $\mathbf{A}_{k}$
on $Y_{kj}$ due to unobserved group properties.

%f16 #&#
%

We also may be interested in whether the effect of an allocation is
mediated by group attributes $\mathbf{T}_{k}$. In order to identify
mediated effects through $\mathbf{T}_{k}$ it must be hypothetically
possible to intervene on $\mathbf{T}_{k}$ without manipulating
any other variable that may cause $Y_{ki}$ not through $\mathbf{T}_{k}$.
For example, if $T_{ki}$ is the enthusiasm of the crowd when $i$'s
group runs the race, then we can imagine intervening on $\mathbf{T}_{k}$
by changing the composition of the crowd without changing the assignments
of runners to heats or their covariates. On the other hand, if $T_{ki}$
is the number of runners in $i$'s group, then clearly any intervention
on $\mathbf{T}_{k}$ must operate because of $\mathbf{A}_{k}$,
which causes $Y_{ki}$ not through $\mathbf{T}_{k}$. Natural
direct, natural indirect and controlled direct effects are not coherently
defined in this case. This is because counterfactuals of the form
$Y_{ki} (\mathbf{a}_{k},\mathbf{T}_{k}(\mathbf{a}_{k}') )$
are not well-defined if we cannot hypothetically simultaneously intervene
on $\mathbf{A}_{k}$, setting it to $\mathbf{a}_{k}$,
and on $\mathbf{T}_{k}$, setting it to its counterfactual value
under allocation $\mathbf{a}_{k}'$. If the only way to set $\mathbf{T}_{k}$
to its counterfactual value under allocation $\mathbf{a}_{k}'$
is through an intervention that sets $\mathbf{A}_{k}$ to $\mathbf{a}_{k}'$,
then this hypothetical joint intervention is not possible. The effects
of $\mathbf{A}_{k}$ on $Y_{ki}$ with $Z_{ki}^{*}$ as a mediator
are similarly incoherent, because it is impossible to imagine intervening
on $Z_{ki}^{*}$ without manipulating $\mathbf{A}_{k}$, $\mathbf{Z}_{k}$,
or both. If we are interested in the role that $Z_{ki}^{*}$ plays
in the effect of group allocation on the outcome, we can instead estimate
the effects of $Z_{ki}^{*}$ on $Y_{ki}$.

%s5 #&#
\section{A Note on Interference and Social Networks}\label{s5}

Our discussion thus far has focused on settings in which individuals
are clustered into blocks and in which individuals in distinct blocks
do not influence each other. In some contexts, it may be the case that
there are no or few distinct independent blocks; social networks constitute
one such setting. A social network is a collection of individuals
and the social ties between them, for example ties of friendship,
kinship or physical proximity. Social networks are of public health
interest because certain health-related behaviors, beliefs and outcomes
may propagate socially (\citeauthor{christakis2007spread},
\citeyear{christakis2007spread,christakis2008collective}; \citeauthor
{smith2008social}, \citeyear{smith2008social}),
but of course they are rife with interference, making causal inference
difficult.

Allocational interference essentially involves intervening on the
network structure itself, creating new ties by assigning individuals
to the same group, for example, assigning children to the same classroom,
and possibly breaking old ties by assigning individuals to different
groups. An intervention on classroom assignments within a school could
be seen as creating a new network topology at the beginning of every
school year. Because the network itself is manipulated in allocational
interference, it may be a useful lens through which to understand
interventions on ties in social network contexts.

\begin{figure}[t]

\includegraphics{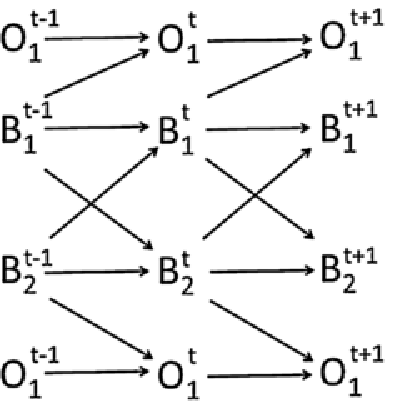}

\caption{}\label{f17}
\end{figure}

\begin{figure}[b]

\includegraphics{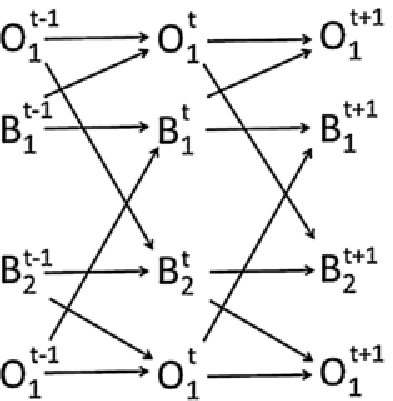}

\caption{}\label{f18}
\end{figure}

Contagion and direct interference occur naturally and widely in social
networks. Direct interference may be present whenever an exposure
consists of ideas, beliefs, knowledge or physical goods which can
be shared by an exposed individual with his associates. Contagion
in social networks has been written about extensively in recent years
(\citeauthor{christakis2007spread},
\citeyear{christakis2007spread,christakis2008collective};
\citeauthor{mulvaney2009obesity}, \citeyear{mulvaney2009obesity};
Smith and Christakis, \citeyear{smith2008social}).
Infectious diseases are more likely to spread between people closely
connected in a social network (e.g., because they live together
or spend time together), and in addition there is some recent evidence
that traits and behaviors like obesity and smoking may be ``socially
contagious'' (\citeauthor{christakis2007spread},
\citeyear{christakis2007spread,christakis2008collective}).
The precise mechanisms for these purported phenomena are unknown,
but some researchers have hypothesized that latent outcomes related
to the observed outcomes may be transmitted through social contact.
For example, \citeauthor{christakis2007spread} (\citeyear
{christakis2007spread,christakis2008collective})
have suggested that beliefs about the acceptability of different smoking
behaviors and body types may be contagious. If this is in fact the
mechanism underlying what appears to be contagion of smoking behavior
or obesity, then the true structure is depicted by the DAG in
Figure~\ref{f17}. $O_{i}^{t}$ represents the observed characteristic
of subject
$i$ at time $t$, for example his smoking behavior, and $B_{i}^{t}$
represents his beliefs, for example, about the acceptability of smoking.
We observe a phenomenon that resembles contagion, namely that $O_{i}^{t}$
appears to have a causal effect on $O_{j}^{t+1}$. However, this apparent
effect may not be due to a causal pathway but rather to the backdoor
path $O_{i}^{t}\leftarrow B_{i}^{t-1}\rightarrow B_{j}^{t}\rightarrow
O_{j}^{t+1}$
as in Figure~\ref{f17}. Another possible structure that would give
rise to
apparent contagion is presented in Figure~\ref{f18}. Here, $O$ is indeed
contagious, but the path by which contagion operates is mediated by
$B$. Distinguishing between these different structures could have
implications for interventions and policies. If the DAG in Figure~\ref
{f17} represents the true causal structure, then intervening on $O$
or introducing a policy targeted at affecting $O$ will not disrupt
the contagious process; we should attempt to intervene on underlying
beliefs instead. If the DAG in Figure~\ref{f18} captures the true mechanisms
at work, then intervening on either $O$ or $B$ can disrupt the contagious
process.

%f17 #&#
%

%f18 #&#
%

The discussion of contagion and direct interference in Section~\ref
{s4} may
be useful for clarifying aspects of social network research. In many
types of social networks and for many types of exposures and outcomes,
both contagion and direct interference will be present. The results
in Section~\ref{s4.2.1} can sometimes be used to differentiate between the
two types of interference. Contagion cannot be identified by cross-sectional
network data without very strong assumptions about temporal and causal
relationships, and some effects related to contagion require fine-grained
information on outcomes in the network over time. Conversely, social
network data can be used to refine assumptions about the structure
of interference. We have assumed that interference occurred between
all individuals in a block. This corresponds to a network in which
each individual has a tie to every other individual in the same block.
But if anything is known about the actual network topology, specifically
about the absence of ties between certain individuals, then this information
could be used to refine the causal structure of interference represented
on the DAGs given in Section~\ref{s3} and, therefore, the conditions under
which causal effects are identifiable.

%s6 #&#
\section{Conclusion}\label{s6}

It is of paramount importance to carefully consider the specific causal
structure whenever interference may operate on the relation between
one individual's treatment and another's outcome. The possible structures
are numerous and valid causal inference may require different assumptions
in each one, depending on the effect of interest and the nature of
confounding.

Also of great importance is the definition of the variables involved
in the causal pathways under investigation. In some cases, the difference
between interference by contagion and direct interference is contextual:
depending on how we define the treatment and the outcome, some causal
relationships can be seen as either one. Recall the example of direct
interference that we presented in Section~\ref{s4.1}: the outcome is weight
change and the treatment dietary counseling from a nutritionist. Direct
interference occurs when a treated individual ``treats'' his associates
by imparting to them the information gained from the nutritionist,
thereby directly affecting their obesity status. Underlying this direct
interference is a contagious process by which the treated individual
transmits his understanding of how to adopt and maintain a healthy
diet to his associates; ``catching'' this understanding causes the
associates to lose weight. Defining variables precisely and narrowly
is always a difficulty for causal inference; the challenge is to conduct
valid inference when we do not observe the underlying causal mechanisms
but instead have to base our analyses on constructs like weight, symptomatic
flu, scholastic achievement, visits with a nutritionist, etc.

This paper scratches the surface of the enormous challenge of causal
inference in the presence of interference. We have not, for example,
touched upon estimation of causal effects or on inference, areas where
some progress has been made in recent years (\citeauthor
{aronow2012estimating}, \citeyear{aronow2012estimating};
\citeauthor{bowers2013reasoning}, \citeyear{bowers2013reasoning};
Graham, Imbens and Ridder, \citeyear{graham2010measuring};
Hudgens and Halloran, \citeyear{hudgens2008toward};
\citeauthor{manski2010identification}, \citeyear{manski2010identification};
\citeauthor{rosenbaum2007interference}, \citeyear
{rosenbaum2007interference}; Tchetgen Tchetgen and VanderWeele,
\citeyear{tchetgen2010causal})
but much more is needed.

%sA #&#
%
\begin{appendix}\label{app}
\section*{Appendix}
We describe the identification of the effects of $A_i$ on $Y_j$ for the
DAGs in Figure~\ref{f5} when $\mathbf{C}$ is not fully observed.

In Figure~\ref{f5}(c), standardizing by $C_i$ identifies the effect of
$A_i$ on $Y_j$:
\begin{eqnarray*}
& & \!\!\sum_{c_{i}}E[Y_{j}|A_{i}=a_{i},C_{i}=c_{i}]P(C_{i}=c)
\\[-1pt]
&& \!\!\quad= \sum_{a_{j}}\sum_{c_{j}}
\sum_{c_{i}}E[Y_{j}|A_{i}=a_{i},
\\[-1pt]
&& \!\!\hphantom{\quad= \sum_{a_{j}}\sum_{c_{j}}
\sum_{c_{i}}E[}{}C_{i}=c_{i},A_{j}=a_{j},C_{j}=c_{j}]
\\
&& \!\!\hphantom{\quad= \sum_{a_{j}}\sum_{c_{j}}
\sum_{c_{i}}}{}\cdot P(A_{j}=a_{j},
\\
&&\!\!\hphantom{\quad= \sum_{a_{j}}\sum_{c_{j}}
\sum_{c_{i}}{}\cdot P(}{}C_{j}=c_{j}|A_{i}=a_{i},C_{i}=c_{i})\\
&& \!\!\hphantom{\quad= \sum_{a_{j}}\sum_{c_{j}}
\sum_{c_{i}}}{}\cdot P(C_{i}=c_{i})
\\
&&\!\!\quad= \sum_{a_{j}}\sum_{c_{j}}
\sum_{c_{i}}E\bigl[Y_{j}(a_{i},a_{j})|A_{i}=a_{i},\\
&& \!\!\hphantom{\quad= \sum_{a_{j}}\sum_{c_{j}}
\sum_{c_{i}}E\bigl[}{} C_{i}=c_{i},A_{j}=a_{j},C_{j}=c_{j}
\bigr]
\\[-1pt]
&& \!\!\hphantom{\quad= \sum_{a_{j}}\sum_{c_{j}}
\sum_{c_{i}}}{}\cdot
P(A_{j}=a_{j},\\
&&\!\!\hphantom{\quad= \sum_{a_{j}}\sum_{c_{j}}
\sum_{c_{i}}{}\cdot P(}{}C_{j}=c_{j}|A_{i}=a_{i},C_{i}=c_{i})
\\
&& \!\!\hphantom{\quad= \sum_{a_{j}}\sum_{c_{j}}
\sum_{c_{i}}}{}\cdot
P(C_{i}=c_{i})
\\
&&\quad= \sum_{a_{j}}\sum_{c_{j}}
\sum_{c_{i}}E\bigl[Y_{j}(a_{i},a_{j})|C_{i}=c_{i},C_{j}=c_{j}
\bigr]\\
&&\!\!\hphantom{\quad= \sum_{a_{j}}\sum_{c_{j}}
\sum_{c_{i}}}{}\cdot
P(A_{j}=a_{j},\\
&&\!\!\hphantom{\quad= \sum_{a_{j}}\sum_{c_{j}}
\sum_{c_{i}}{}\cdot P(}{}C_{j}=c_{j}|A_{i}=a_{i},C_{i}=c_{i})\\
&&\!\!\hphantom{\quad= \sum_{a_{j}}\sum_{c_{j}}
\sum_{c_{i}}}{}\cdot P(C_{i}=c_{i}).
\end{eqnarray*}
This is a weighted average of $\mathbf{C}$-specific counterfactuals
$Y_{j}(\mathbf{a})$. [Replacing $C_{i}$ with $D$ in the expressions
above gives the identifying expression for the effect of $A_{i}$
on $Y_{j}$ in Figure~\ref{f5}(e).] Standardizing by $C_{i}$ also
identifies the effect of $A_{i}$
on $Y_{i}$ for the DAGs in Figures~\ref{f5}(a) and~\ref{f5}(c),
similarly giving
a weighted average of \mbox{$\mathbf{C}$-specific} counterfactuals:
\begin{eqnarray*}
& &\!\! \sum_{c_{i}}E[Y_{i}|A_{i}=a_{i},C_{i}=c_{i}]P(C_{i}=c)
\\
&&\!\!\quad= \sum_{a_{j}}\sum_{c_{j}}
\sum_{c_{i}}E[Y_{i}|A_{i}=a_{i},
\\
&&\!\! \hphantom{\quad= \sum_{a_{j}}\sum_{c_{j}}
\sum_{c_{i}}E[}{}C_{i}=c_{i},A_{j}=a_{j},C_{j}=c_{j}]
\\
&&\!\!\hphantom{\quad= \sum_{a_{j}}\sum_{c_{j}}
\sum_{c_{i}}}{}\cdot P(A_{j}=a_{j},\\
&&\!\!\hphantom{\quad= \sum_{a_{j}}\sum_{c_{j}}
\sum_{c_{i}}{}\cdot P(}{}C_{j}=c_{j}|A_{i}=a_{i},C_{i}=c_{i})\\
&&\!\!\hphantom{\quad= \sum_{a_{j}}\sum_{c_{j}}
\sum_{c_{i}}}{}\cdot P(C_{i}=c_{i})
\\
&&\!\!\quad= \sum_{a_{j}}\sum_{c_{j}}
\sum_{c_{i}}E\bigl[Y_{i}(a_{i},a_{j})|A_{i}=a_{i},
\\
&&\!\!\hphantom{\quad= \sum_{a_{j}}\sum_{c_{j}}
\sum_{c_{i}}E\bigl[}{} C_{i}=c_{i},A_{j}=a_{j},C_{j}=c_{j}
\bigr]\\
&&\!\!\hphantom{\quad= \sum_{a_{j}}\sum_{c_{j}}
\sum_{c_{i}}}{}\cdot P(A_{j}=a_{j},\\
&&\!\!\hphantom{\quad= \sum_{a_{j}}\sum_{c_{j}}
\sum_{c_{i}}{}\cdot P(}{}C_{j}=c_{j}|A_{i}=a_{i},C_{i}=c_{i})
\\
&&\!\!\hphantom{\quad= \sum_{a_{j}}\sum_{c_{j}}
\sum_{c_{i}}}{}\cdot P(C_{i}=c_{i})
\\
&&\!\!\quad= \sum_{a_{j}}\sum_{c_{j}}
\sum_{c_{i}}E\bigl[Y_{i}(a_{i},a_{j})|C_{i}=c_{i},C_{j}=c_{j}
\bigr]
\\
&&\!\! \hphantom{\quad= \sum_{a_{j}}\sum_{c_{j}}
\sum_{c_{i}}}{}\cdot P(A_{j}=a_{j},\\
&&\!\!\hphantom{\quad= \sum_{a_{j}}\sum_{c_{j}}
\sum_{c_{i}}{}\cdot P(}{}C_{j}=c_{j}|A_{i}=a_{i},C_{i}=c_{i})
\\
&&\!\!\hphantom{\quad= \sum_{a_{j}}\sum_{c_{j}}
\sum_{c_{i}}}{}\cdot P(C_{i}=c_{i}).
\end{eqnarray*}

\end{appendix}

\section*{Acknowledgment}
This work was supported by NIH Grant ES017876.

%suskaldyti doi

% imsref loaded by arune.pranskunaite, 2014-10-21 12:24:24
%

%
\end{document}